\documentclass[10pt,conference]{IEEEtran}
\usepackage{cite}
\usepackage{amsmath,amssymb,amsfonts}
\usepackage{algorithmic}
\usepackage{graphicx}
\usepackage{textcomp}
\usepackage[dvipsnames]{xcolor}
\usepackage{xspace}
\usepackage{xfrac}
\usepackage{balance}
\usepackage[all]{nowidow}
\usepackage{kotex}
\usepackage{lipsum}
\usepackage{enumitem}
\usepackage[hyphens]{url}
\usepackage{fancyhdr}
\usepackage{hyperref}
\usepackage{booktabs}
\usepackage{array}
\usepackage{hhline}
\usepackage{makecell}
\usepackage{multirow}
\usepackage[utf8]{inputenc}
\usepackage{blindtext}
\usepackage{tikz}
\usepackage{longtable}
\usepackage{comment}
\usepackage[table]{xcolor}
\usepackage{mathtools}
\usepackage{tcolorbox}

\def\BibTeX{{\rm B\kern-.05em{\sc i\kern-.025em b}\kern-.08em
    T\kern-.1667em\lower.7ex\hbox{E}\kern-.125emX}}

\newcommand{\dev}{accelerator\xspace}
\newcommand{\devs}{accelerators\xspace}

\newcommand{\gpub}{\textbf{256 GPU}\xspace}
\newcommand{\gpus}{\textbf{32 GPU$\times$8}\xspace}

\newcommand{\ai}{ArI\xspace}
\newcommand{\ie}{\emph{i.e.}\xspace}
\newcommand{\eg}{\emph{e.g.}\xspace}

\newcommand{\ti}[1]{$\mathbf{#1}_{i}$}
\newcommand{\tut}[2]{${#1}_\mathrm{#2}$}
\definecolor{lightergray}{gray}{0.95} 

\newcommand*{\Scale}[2][4]{\scalebox{#1}{$#2$}}
\newcommand{\niparagraph}[1]{\noindent \textbf{#1}}

\title{Rethinking LLM Inference Bottlenecks: Insights from Latent Attention and Mixture-of-Experts}

\author{\IEEEauthorblockN{
Sungmin Yun\IEEEauthorrefmark{2},
Seonyong Park\IEEEauthorrefmark{2},
Hwayong Nam\IEEEauthorrefmark{2},
Younjoo Lee\IEEEauthorrefmark{2},
Gunjun Lee\IEEEauthorrefmark{2},
Kwanhee Kyung\IEEEauthorrefmark{2},\\
Sangpyo Kim\IEEEauthorrefmark{4},
Nam Sung Kim\IEEEauthorrefmark{3},
Jongmin Kim\IEEEauthorrefmark{2},
Hyungyo Kim\IEEEauthorrefmark{3},
Juhwan Cho\IEEEauthorrefmark{2},
Seungmin Baek\IEEEauthorrefmark{2},
Jung Ho Ahn\IEEEauthorrefmark{2}
}

\IEEEauthorblockA{
\IEEEauthorrefmark{2}\textit{Seoul National University, Seoul, South Korea}, \\
\IEEEauthorrefmark{3}\textit{University of Illinois at Urbana-Champaign, Champaign, Illinois, USA} \\
\IEEEauthorrefmark{4}\textit{CryptoLab Inc., Seoul, South Korea} \\
\textit{\{sungmin.yun, seonyong.park, nhy4916, younjoo0614, kevin970401, kwanhee5,} \\
\textit{vnb987, jongmin.kim, jfcho2, qortmdalss, gajh\}@snu.ac.kr} \\
\textit{\{spkim\}@cryptolab.co.kr} \\
\textit{\{hyungyo2, nskim\}@illinois.edu}
}
}

\begin{document}
\pagenumbering{arabic}
\maketitle
\thispagestyle{plain}
\pagestyle{plain}

\begin{abstract}
Computational workloads composing traditional transformer models are starkly bifurcated.
Multi-Head Attention (MHA) and Grouped-Query Attention are memory-bound due to low arithmetic intensity, while FeedForward Networks are compute-bound.
This dichotomy has long motivated research into specialized hardware to mitigate the attention bottleneck.

This paper argues that recent architectural advances in transformer models---Multi-head Latent Attention (MLA) and Mixture of Experts (MoE)---introduce new dominant bottlenecks, shifting the challenge away from memory-intensive attention.
We make two key observations.
First, the arithmetic intensity of MLA is over two orders of magnitude higher than that of MHA, moving it toward a compute-bound regime well-matched to modern accelerators such as GPUs.
Second, distributing MoE experts across a pool of accelerators allows batching to tune their arithmetic intensity to that of dense layers, producing a more balanced computational profile.
Consequently, the focus of hardware and system optimization should shift from attention acceleration to high-bandwidth interconnects and balancing expert workloads across accelerators.

\end{abstract}

\section{Introduction}
\label{sec:1_introduction}
Transformer-based large language models (LLMs)~\cite{neurips_2017_transformer} have achieved remarkable accuracy across various natural language processing tasks~\cite{arxiv-2025-deepseek-r1}.
An LLM summarizes (or prefills) an input sequence of tokens; then, it generates output tokens one-by-one through decode steps.
A conventional LLM consists of a sequence of decoder blocks, each comprising a Multi-Head Attention (MHA) or Grouped-Query Attention (GQA) sub-block and a FeedForward Network (FFN) sub-block.

To improve scalability and efficiency, recent LLM architectures have adopted optimizations such as Multi-head Latent Attention (MLA)~\cite{arxiv-2025-deepseek-r1}, first appeared in DeepSeek-V2~\cite{arxiv-2024-deepseek-v2}, which reduces the memory footprint of attention.
They also incorporate Mixture of Experts (MoE)~\cite{acl-2024-deepseek-moe, jmlr-2022-switchtransformer, github-2024-grok1, arxiv-2024-mixtral, meta-2025-llama4} to increase the model capacity with multiple \emph{experts} for the FFN sub-block, where only a subset of experts are activated per token to mitigate the compute cost~\cite{arxiv-2017-outrageouslylargeneuralnetworks}.

Maximizing accelerator utilization when serving these models is critical for improving throughput and end-to-end latency~\cite{osdi-2024-sarathi}. 
Serving LLMs in production requires deploying systems that integrate hundreds or even thousands of accelerators (\eg, GPUs~\cite{nvidia-gb200-nvl72} and TPUs~\cite{google-tpuv7-blog}) to handle high query volumes and large models~\cite{sc-2022-deepspeed-inference, mlsys-2023-scaling}. 
Inefficient utilization results in compute resources remaining idle, higher infrastructure costs, and failure to meet service-level objectives (SLOs)~\cite{osdi-2024-distserve}.

A key factor in serving (inferencing) LLMs efficiently is arithmetic intensity (\ai), the ratio of arithmetic operations to memory access, measured in operations per byte (Op/B).
The ridge point of an accelerator, a concept from the roofline model~\cite{commun-2009-roofline}, defines the \ai at which performance transitions from being \emph{memory-bound} to \emph{compute-bound}.
To fully exploit an accelerator's capabilities, the \ai of each layer in the model should be configured to approach this ridge point.

In this paper, we argue that the introduction of MLA and MoE fundamentally reshapes the computational landscape of LLM inference.
Crucially, both techniques significantly reduce per-request memory capacity and computation demands, especially in the decode stage.
By introducing a latent space to attention, MLA significantly reduces the KV cache (KV\$) size, a major bottleneck in conventional LLMs.
The reduced KV\$ size enables the use of much larger batch sizes.
Furthermore, while MoE significantly increases the model size with multiple experts, associated computational costs do not increase proportionally due to the sparse activation of the experts.

We demonstrate how these architectural shifts in LLMs synergistically increase throughput while reducing latency compared to conventional models by operating near the accelerator’s ridge point.
With layer reordering applied in the decode stage of an MLA sub-block, the \ai of its core-attention layer approaches the accelerator’s ridge point regardless of the batch size, by reducing the number of memory accesses.
This enables significantly large batch sizes while still satisfying SLOs.
Along with the reduced per-expert computation, the large batch sizes also drive the \ai of each expert’s FC layers closer to the accelerator’s ridge point.
This, in turn, enables serving systems to support MoE layers more efficiently.

Building on these observations, we present a serving-system methodology that holistically integrates MLA and MoE within multi-accelerator environments.
We distill our findings into three design principles for large-batch inferences, aiming to maximize throughput while satisfying SLO constraints.
First, \emph{attention-specialized hardware such as processing-in-memory (PIM) is no longer necessary} as MLA and MoE shift attention and expert computation toward the compute-bound regime. 
Second, \emph{high-bandwidth interconnects (e.g., NVLinks) are indispensable} for reducing latency in token dispatch and aggregation within MoE layers.
Third, \emph{balancing expert workloads is critical}; mitigating load imbalance across accelerators caused by skewed expert distributions~\cite{neurips-2024-moe-efficient} preserves throughput scalability. 
These principles guide the design of balanced, high-throughput LLM serving systems co-optimized across model, hardware, and interconnect levels.

The key contributions of this paper are as follows:
\begin{itemize}[leftmargin=*,nolistsep]
\item We discover that layer reordering in MLA increases the arithmetic intensity of core-attention layer, making it approach the accelerator's ridge point.

\item We demonstrate that the reduced KV-cache size in MLA and the decreased computation from sparsely activated MoE experts synergistically enable efficient serving of large batches.

\item We highlight that the inference bottlenecks lie in interconnect bandwidth and load imbalance across tokens.

\end{itemize}

\section{Background}
\label{sec:2_background}

\subsection{Standard LLM architecture and its layers}
\label{subsec:2_1_llm}

\begin{figure}[!tb]
\centering
\includegraphics[width=0.98\columnwidth]{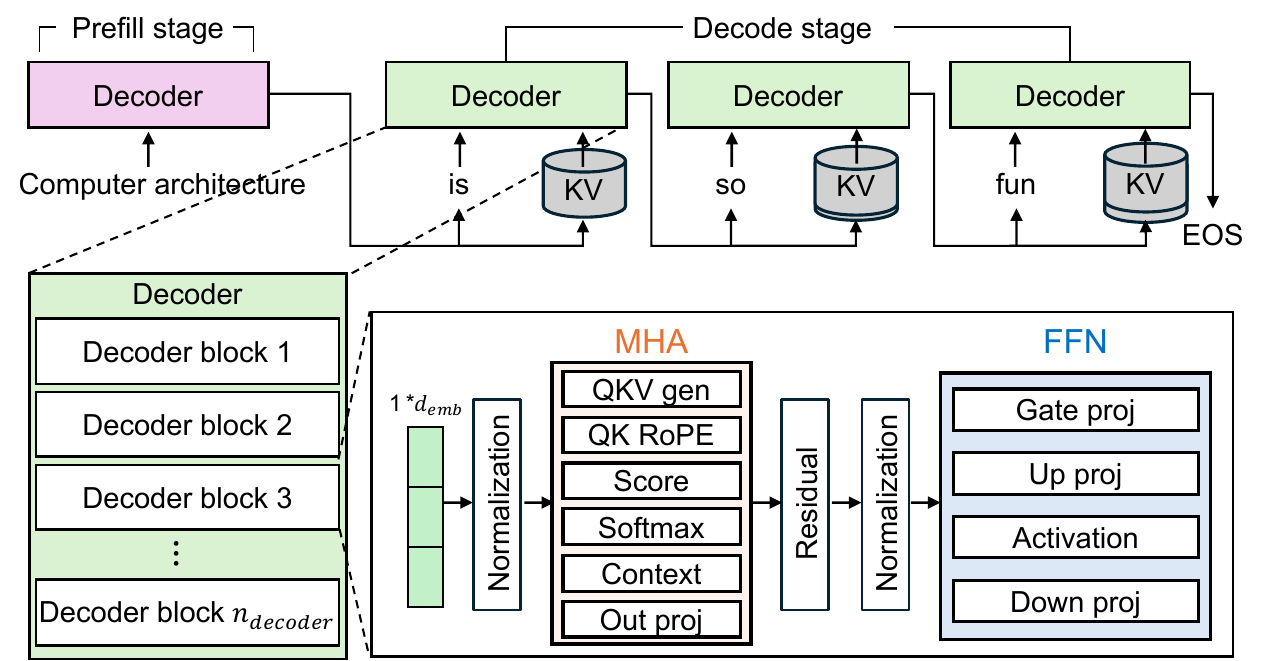}
\vspace{-0.1in}
  \caption{Conventional transformer-decoder-based LLM architecture.}
  \label{fig:llm_arch}
\vspace{-0.1in}
\end{figure}

Despite rapid advancement in LLM algorithms~\cite{arxiv-2024-mamba, arxiv-2025-diffusionllm}, transformer-based architectures~\cite{neurips_2017_transformer}, especially those only with decoders~\cite{neurips-2020-gpt-3}, remain the standard backbone for modern LLMs (Figure~\ref{fig:llm_arch}).
A transformer consists of a series of $n_\mathrm{decoder}$ decoder blocks.
Given an input sequence of $\ell$ tokens (\eg, words) for an inference (serving) request, each token is embedded into a $d_\mathrm{emb}$-dimensional hidden state for $d_\mathrm{emb}$ typically on the order of thousands.
This forms a hidden state matrix $\mathbf{H}_{\ell} \in \mathbb{R}^{\ell \times d_\mathrm{emb}}$ that is provided as input to the decoder blocks.
This matrix passes through the decoder blocks, each with its own trained weights, and is transformed into an output vector.

LLM inference consists of a prefill (summarization) stage and a decode (generation) stage.
The prefill stage processes the entire input hidden state matrix ($\ell = L_\mathrm{in}$) 
to generate the first output token. 
This stage mainly performs matrix-matrix multiplications, implemented as GEMM operations.
Subsequently, the decode stage generates the remaining tokens auto-regressively, where each step takes the previously generated single token ($\ell=1$) as input to produce the next one, continuing until an end-of-sequence token is generated or the output sequence reaches the maximum output token. 
Also, this stage is dominated by GEMV operations, as each step multiplies a single-token vector ($\ell=1$) with weight matrices for a single-batch inference.

A decoder block in a conventional LLM consists of two sub-blocks: a \textbf{Multi-Head Attention} (MHA) sub-block and a \textbf{FeedForward Network} (FFN) sub-block.
In the rest of the paper, we refer to a sub-block simply as a block.
In MHA, a single hidden state matrix $\mathbf{H}$ (hence \emph{self}-attention) is first linearly transformed (projected) into \textbf{\underline{Q}}uery (Q), \textbf{\underline{K}}ey (K), and \textbf{\underline{V}}alue (V) matrices, each dimensioned by a decompression dimension $d_\mathrm{dec}$, by passing through fully-connected (FC) layers with pre-trained weights.
These matrices are split into $n_\mathrm{hd}$ heads, each dimensioned by $d_\mathrm{hd}$ (\ie, $d_\mathrm{dec}=n_\mathrm{hd} \cdot d_\mathrm{hd}$).
Eq.~\ref{eq:qkv_gen} shows how $\mathbf{Q}, \mathbf{K}$, and $\mathbf{V}$ for each head are computed where $L$ denotes the current sequence length, defined as the sum of the input sequence length $L_{in}$ and the number of tokens generated so far.
During auto-regressive decoding, KV cache (KV\$) stores past K and V values to maintain context without costly recomputation. 
Thus, only the new K and V vectors for the current input token are computed and appended to KV\$.

\begin{equation}
\label{eq:qkv_gen}
\Scale[1.0]{
\begin{split}
\underbrace{\mathbf{Q}_{i}}_{\mathbb{R}^{\ell \times \frac{d_{\mathrm{dec}}}{n_{\mathrm{hd}}}}} =\underbrace{\mathbf{H}_{\ell}}_{\mathbb{R}^{\ell \times d_{\mathrm{emb}}}} \cdot \underbrace{\mathbf{W}_{\mathbf{Q}_{i}}}_{\mathbb{R}^{d_{\mathrm{emb}} \times\frac{d_{\mathrm{dec}}}{n_{\mathrm{hd}}}}} \\
\underbrace{\mathbf{X}_{i}}_{\mathbb{R}^{L \times \frac{d_{\mathrm{dec}}}{n_{\mathrm{hd}}}}} =\underbrace{\mathbf{H}_{L}}_{\mathbb{R}^{L \times d_{\mathrm{emb}}}} \cdot \underbrace{\mathbf{W}_{\mathbf{X}_{i}}}_{\mathbb{R}^{d_{\mathrm{emb}} \times\frac{d_{\mathrm{dec}}}{n_{\mathrm{hd}}}}}
\\
\text{for } \mathbf{X} \in \{\mathbf{K}, \mathbf{V}\}  \text{ and } i \in [1, n_{\text{hd}}]
\end{split}
}
\end{equation}
\noindent 
Each head independently performs a sequence of operations referred to as a core-attention layer, 
which computes score (Eq.~\ref{eq:score}), softmax, and context (Eq.~\ref{eq:context}) operations.

\begin{equation}
\label{eq:score}
\Scale[1.0]{
\begin{split}
\underbrace{\mathbf{S}_{i}}_{\mathbb{R}^{\ell \times L}} = \underbrace{\mathbf{Q}_{i}}_{\mathbb{R}^{\ell \times \frac{d_{\text{dec}}}{n_{\text{hd}}}}} \cdot \underbrace{\mathbf{K}_{i}^{\text{T}}}_{\mathbb{R}^{\frac{d_{\text{dec}}}{n_{\text{hd}}}\times L}}
\end{split}}
\end{equation}
\vspace{0.1em}
\begin{equation}
\label{eq:context}
\Scale[1.0]{
\underbrace{\mathbf{O}_{i}}_{\mathbb{R}^{\ell \times \frac{d_{\mathrm{dec}}}{n_{\text{hd}}} }} = \underbrace{\text{Softmax}({\frac{\mathbf{S}_{i}}{\sqrt{d_{\mathrm{dec}}/{n_{\text{hd}}}}}})}_{\mathbb{R}^{\ell \times L}}\cdot\underbrace{\mathbf{V}_{i}}_{\mathbf{R}^{L\times \frac{d_{\mathrm{dec}}}{n_{\text{hd}}}}}
}
\end{equation}
\noindent Finally, another FC layer called attention output projection follows, generating the MHA block's output \textbf{U}.

\textbf{Grouped-Query Attention (GQA)} has been widely adopted in modern LLMs~\cite{github-2024-grok1, arxiv-2024-llama3, meta-2025-llama4, arxiv-2024-mixtral}.
GQA groups multiple Q heads ($deg_\mathrm{grp}$) to share a single KV head within each group.
GQA reduces the KV\$ size at the cost of possible reduction in accuracy~\cite{emnlp-2023-gqa}.
While Multi-Query Attention (MQA)~\cite{arxiv-2019-mqa} further reduces KV\$ size by sharing a single KV head across all Q heads, this significantly degrades accuracy~\cite{emnlp-2023-gqa}. 
Hence, we exclude MQA from our analysis.

An FFN block in recent LLMs consists of three FC layers and one non-linear activation. 
Until GPT-3~\cite{neurips-2019-gpipe}, most models used an FFN block with two FC layers and one activation.
More recent LLMs employ three FC layers, which improves response quality at the cost of additional computation by introducing gated activation functions~\cite{arxiv-2020-glu}.

Modern LLMs commonly apply Rotary Positional Embedding (RoPE)~\cite{arxiv-2023-roformer} to inject positional information into token generations. 
RoPE encodes token positions by applying a position-dependent rotational transformation to Q and K before the core-attention layer while leaving V unchanged.

In LLM inference, multiple requests in a batch share the same model weights, allowing weight reuse and reducing memory access for FC layers.
In the decode stage, batching converts FC-layer operations from GEMV to GEMM, as activations from multiple requests are multiplied jointly with the model weights.
However, as each request accompanies its own KV\$, increasing the batch size $B$ results in higher memory usage. 
As a result, the maximum feasible batch size is constrained by both memory capacity and SLOs~\cite{osdi-2024-distserve}.

\subsection{Hardware efficiency \& arithmetic intensity (\ai)}
\label{sec:roofline}

When serving LLMs, three key factors determine service latency and throughput: \emph{arithmetic throughput}, \emph{memory bandwidth}, and \emph{memory capacity} of an accelerator.
High arithmetic throughput enables fast execution of compute-intensive layers, such as GEMMs on large tensors during the prefill stage (\emph{e.g.}, Eq.~\ref{eq:qkv_gen}). 
To fully leverage this arithmetic capability, sufficient memory bandwidth is essential to quickly supply the data required for these computations.
Such high bandwidth can only be fully utilized when the memory capacity is large enough to accommodate the entire working set; otherwise, data must be fetched from lower-bandwidth sources (\eg, via a PCIe link), severely limiting performance.

\ai is useful for evaluating the expected throughput of an algorithm with ample parallelism on a given accelerator (e.g., GPU or TPU).
It is the ratio of operations performed to the amount of data accessed from memory in bytes (Op/B). 
The \emph{ridge point} of an accelerator ($\mathrm{RP_{acc}}$)~\cite{commun-2009-roofline} refers to the \ai at which performance shifts from being memory-bound to compute-bound.
It is calculated as the ratio of peak arithmetic throughput to peak memory bandwidth of the accelerator.
If the \ai of a computation is below the ridge point, its performance is limited by memory bandwidth; its execution time is largely determined by $\frac{\text{\# of memory accesses}}{\text{memory bandwidth}}$.
If it is above $\mathrm{RP_{acc}}$ such that a sufficient number of operations are performed per unit of data to make the computation compute-bound, the achieved arithmetic throughput will saturate.
Then, the execution time would be bounded by $\frac{\text{\# of operations}}{\text{arithmetic throughput}}$~\cite{commun-2009-roofline}.

\subsection{Model parallelism}
\label{sec:background_serving_system}
To distribute model weights and computations across multiple \devs~\cite{blog-2024-nvidia-1, arxiv-2025-genz}, modern LLM serving systems employ tensor parallelism (TP) and data parallelism (DP)~\cite{arxiv-2019-megatron-lm}.
With model size now exceeding the capacity of a single accelerator device, multi-\dev serving has become essential.
TP partitions activation, weights, or both across \devs, allowing each to compute partial results of a single operation in parallel; however, TP introduces inter-\dev communication.
DP replicates weights and processes independent sub-batches on each \dev to avoid the communication but requires larger memory capacity.
We use $deg_\mathrm{TP}$ and $deg_\mathrm{DP}$ to denote the degrees of TP and DP, respectively.

\section{Emerging LLM Optimizations: Impact on Performance and Memory Capacity Demand}
\label{sec:optimizations}
\begin{figure}[!t]
\centering
\includegraphics[width=0.99\columnwidth]{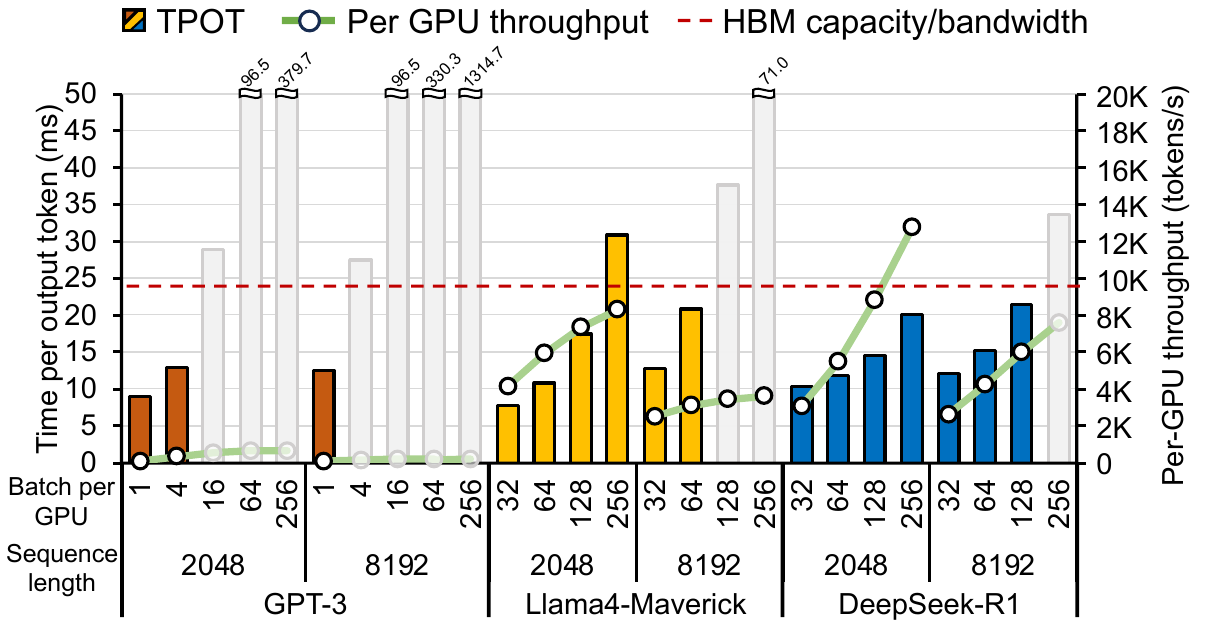}
  \vspace{-0.1in}
  \caption{Time per output token (TPOT) and per-GPU throughput of GPT-3, Llama4-Maverick, and DeepSeek-R1 across various sequence lengths and batch sizes. 
  The blurred area indicates configurations where out-of-memory errors occur.
  See \S\ref{sec:experimental_setup} for experimental details.}
  \label{fig:GPT_DeepSeek_TPOT_throughput}
  \vspace{-0.08in}
\end{figure}

For efficient LLM serving, a myriad of optimizations have been proposed, spanning both algorithmic strategies and hardware-aware techniques, in response to ever-growing demand for LLM services.
Contrary to conventional LLMs, DeepSeek-R1~\cite{arxiv-2025-deepseek-r1} leverages \textbf{Multi-head Latent Attention (MLA)} and \textbf{Mixture of Experts (MoE)}.
Figure~\ref{fig:GPT_DeepSeek_TPOT_throughput} shows that DeepSeek-R1 delivers up to 41$\times$ and 2$\times$ higher per-device throughput, and 30\% and 34\% lower time per output token (TPOT) than GPT-3 and Llama4-Maverick, respectively, even with 3.8$\times$ and 1.7$\times$ larger model capacity (number of parameters).
We provide a high-level explanation of two key \textbf{\underline{R}}easons why DeepSeek-R1 offers higher throughput and lower latency compared to conventional LLMs.

\setlength{\tabcolsep}{2pt}

\begin{table*}[!htb]
  \centering
  \caption{Arithmetic throughput, main-memory bandwidth and capacity of deep-learning accelerators}
  \vspace{-0.1in}
  \resizebox{0.95\textwidth}{!}{%
  \begin{tabular}{l|c|c|c|c|c|c|c}
    \toprule
    & V100 SXM2~\cite{nvidia-v100} & A100 SXM4~\cite{nvidia-a100} & H200 SXM5~\cite{nvidia-h100} & B200 SXM6~\cite{nvidia-b200} & TPU~V5P~\cite{google-tpuv5p-blog} & TPU~V7~\cite{google-tpuv7-blog} & MI325X~\cite{amd-gpumi325x-datasheet}\\
    \midrule
    BF16 throughput (TFLOPS)            & 125   & 312   & 989.5 & 2250 & 459 & 2307 & 1307.4 \\
    \hline
    Memory bandwidth (GB/s)                     & 900   & 2039  & 4800  & 8000 & 2765 & 7400 & 6000 \\
    \hline
    HBM capacity per GPU (GB)                   & 32    & 80    & 141   & 192  & 95  & 192 & 256 \\
    \hline
    Ridge point (BF16)   & 138.89& 153.02& 206.15& 281.25& 166 & 320.42 & 217.9\\
    \bottomrule
  \end{tabular}
  \vspace{-0.2in}
  }%
  \label{tab:ai_accelerator}
\end{table*}

\begin{figure}[!bt]
\centering
\includegraphics[width=0.99\columnwidth]{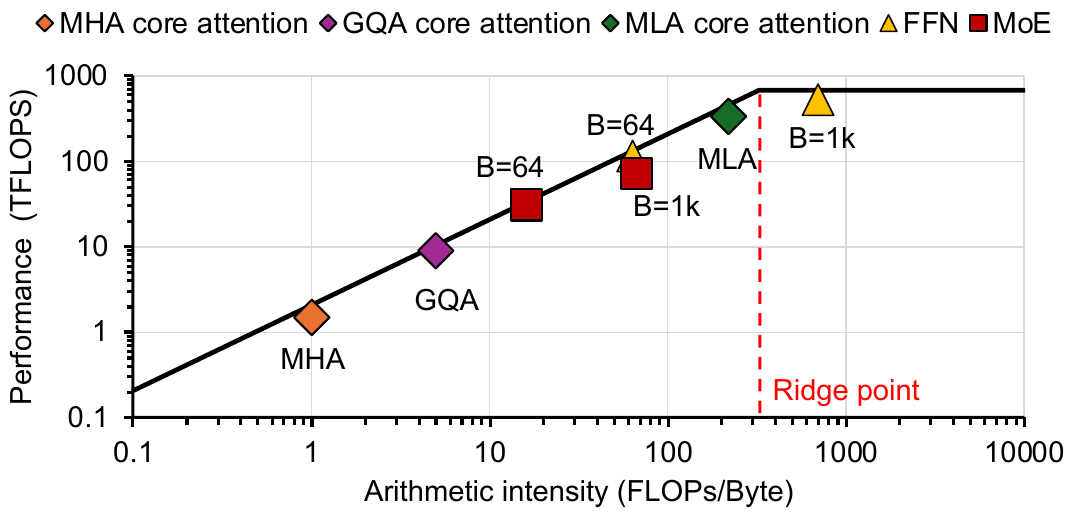}
  \vspace{-0.1in}
  \caption{
  Roofline plot of layers in decoder block at sequence length $L=4096$, obtained from real-machine measurements on an NVIDIA H100 GPU.}
  \label{fig:roofline}
  \vspace{-0.08in}
\end{figure}

\niparagraph{(R1) Reduced core-attention layer latency in MLA:}
Figure~\ref{fig:roofline} shows that the core-attention layer of GPT-3, based on MHA, presents an \ai of approximately 1 even with batching, making it the primary bottleneck preventing throughput and latency improvements during the decode stages. 
Unlike FC layers, which amortize the cost of accessing their weights by reusing them across multiple batched requests, the core-attention layer operates on KV\$ and attention score values, which are unique to each request. 
Thus, these values cannot be shared across batched requests, limiting \ai to 1 regardless of $B$ in MHA.
Because of this extremely low \ai, performance is bounded by memory bandwidth, leading to severe underutilization of compute resources.
GQA reduces the number of memory accesses by sharing KV\$ across multiple queries, thereby slightly improving the \ai; in Llama4-Maverick, each KV\$ is shared among five queries.
Still, the operations remain largely bounded by memory bandwidth.

By contrast, the core-attention layer of DeepSeek-R1, based on MLA, significantly reduces the number of memory accesses and brings the \ai close to the ridge point of modern accelerators ($\mathrm{RP_{acc}}$) through layer reordering (detailed in \S\ref{sec:contribution-2}) as shown in Figure~\ref{fig:roofline}.
This not only reduces the core-attention layer latency but also increases compute utilization during the decode stages.

\niparagraph{(R2) Larger \textit{B} and less computation for MoE:} 
\noindent Larger batch size ($B$) improves the \ai of memory-bound FC layers, enhancing compute utilization of \devs and maximizing throughput with negligible latency increase.
However, $B$ is limited by the KV\$ size, which grows with sequence length.
In GPT-3, the KV\$ size of a request reaches 9 GB for a sequence length of 2048 (detailed in \S\ref{sec:contribution-4}).
The large KV\$ size limits $B$, leaving FC layers memory-bound, reducing throughput and underutilizing compute resources~\cite{micro-2024-duplex, asplos-2024-neupim}.

\begin{figure}[!bt]
\centering
\includegraphics[width=0.99\columnwidth]{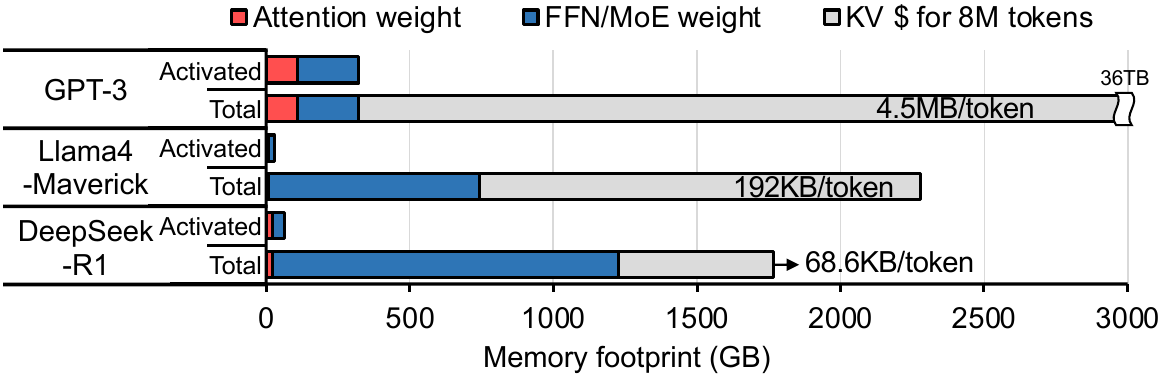}
  \vspace{-0.1in}
  \caption{Comparison of GPT-3, Llama4-Maverick, and DeepSeek-R1 for 8M tokens with BF16 precision: the amount of parameters accessed and computed per token (Activated) and the total memory capacity required for storing attention weights, FFN/MoE weights, and KV\$  (Total).}
  \label{fig:GPT_DeepSeek_mem_usage}
  \vspace{-0.1in}
\end{figure}

In contrast, DeepSeek-R1 requires more memory to store all model parameters, but it uses 67$\times$ smaller KV\$ per token compared to GPT-3, as shown in Figure~\ref{fig:GPT_DeepSeek_mem_usage}.
This reduction enables larger $B$ and higher throughput under the same memory budget.
Also, while GPT-3 performs computations with all 175B parameters for every token, DeepSeek-R1, leveraging MoE, activates only 8 out of 256 experts per token.
As a result, it utilizes just 37B out of 671B total parameters.
This sparsity significantly reduces the per-token computation cost, making DeepSeek-R1 more suitable for large-batch inference.

Llama4-Maverick uses 2.84$\times$ larger KV\$ per token than Deepseek-R1.
Given that the hidden dimension of Llama4-Maverick is 5120, whereas that of DeepSeek-R1 is 7168, we observe that MLA effectively reduces the KV\$ size compared to GQA—by approximately 4$\times$ for the same hidden dimension.
Consequently, for a sequence length of 8192, Llama4-Maverick cannot support a batch size of 128 per GPU, whereas DeepSeek-R1 can, despite its larger model size (see Fig.~\ref{fig:GPT_DeepSeek_TPOT_throughput}).

Lastly, further increasing $B$ drives the \ai beyond the $\mathrm{RP_{acc}}$ and makes the FC layers compute-bound, hence increasing latency without a corresponding throughput improvement. 
Thus, any changes to an algorithm and/or its operating parameters, such as $B$, must consider the $\mathrm{RP_{acc}}$ of the target serving system.
The latest B200 GPU, for instance, provides approximately 18$\times$ higher arithmetic throughput than a V100 GPU, representing the most significant improvement among the accelerators listed in Table~\ref{tab:ai_accelerator}.
Nevertheless, since both arithmetic throughput and memory bandwidth have scaled with the technology, the B200's $\mathrm{RP_{acc}}$ increases by only a modest factor of 2 compared to the V100's;
contemporary accelerators exhibit $\mathrm{RP_{acc}}$s within a narrow range of 200--400 Op/B.

\begin{figure*}[!htb]
  \center
  \includegraphics[width=0.99\textwidth]{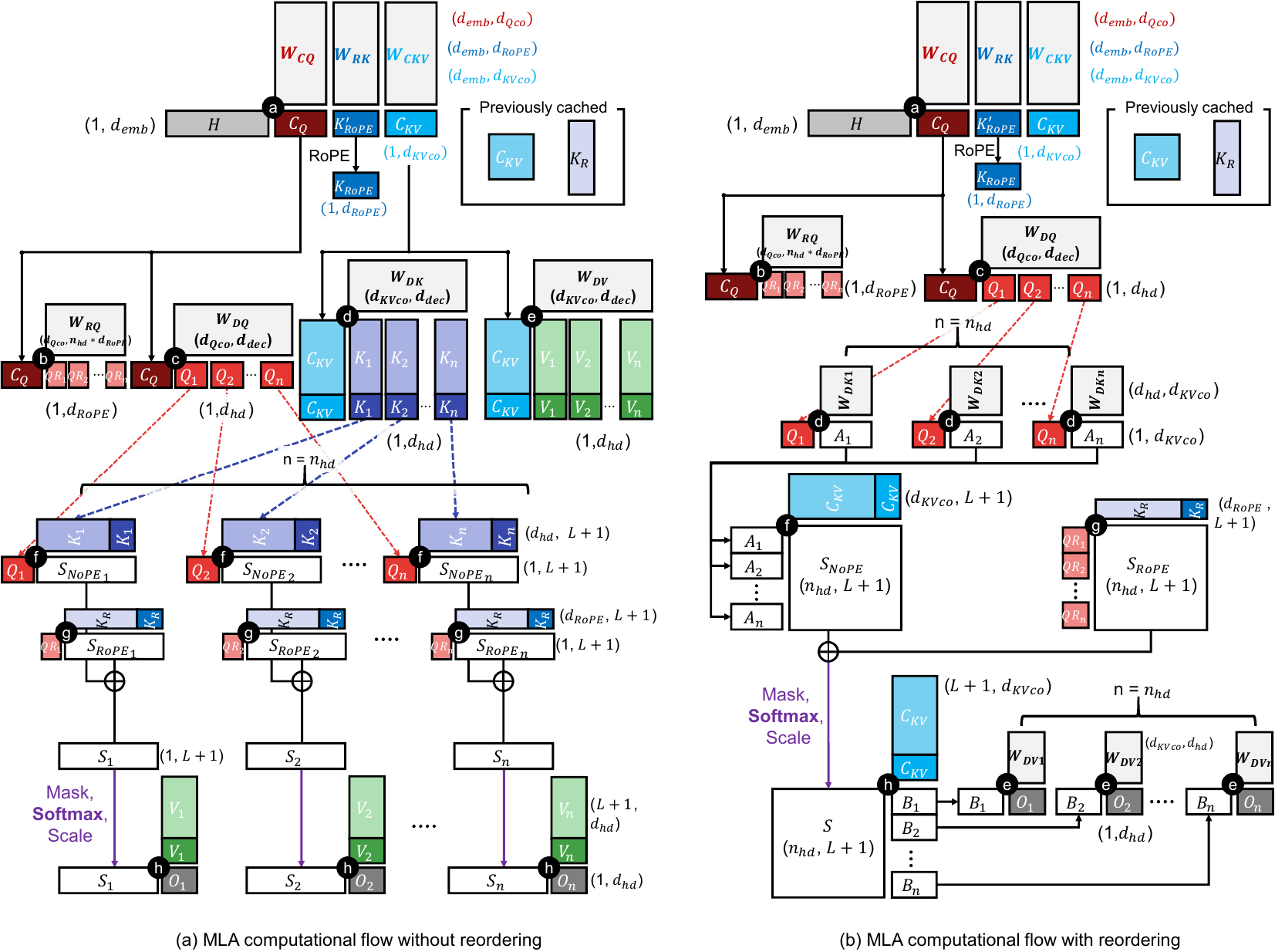}
  \vspace{-0.0in}
  \caption{Computation flow of multi-head latent attention (MLA) with/without layer reordering.
  \textcircled{\scriptsize{a}} to \textcircled{\scriptsize{h}} refer to the layers of MLA (e.g., \textcircled{\scriptsize{a}}: QKV compression, \textcircled{\scriptsize{b}}: Q RoPE, \textcircled{\scriptsize{c}}: Q decompression, \textcircled{\scriptsize{d}}: K decompression, \textcircled{\scriptsize{e}}: V decompression, \textcircled{\scriptsize{f}}: Score, \textcircled{\scriptsize{g}}: K RoPE, \textcircled{\scriptsize{h}}: Context).
  }
  \label{fig:mlaflow_topdown}
  \vspace{-0.1in}
\end{figure*}

\section{Insights on Multi-head Latent Attention}
\label{sec:contribution-2}

\begin{table}[!tb]
\caption{Symbols used throughout this paper, their descriptions, and the exemplar parameters used in DeepSeek-R1~\cite{arxiv-2025-deepseek-r1}}
\vspace{-0.1in}
\centering
\label{tab:deepseek_r1_symbol}
\begin{tabular} {l|l|l}
\toprule
\textbf{Term} & \textbf{Description} & \textbf{DeepSeek-R1} \\
\midrule
$d_{\mathrm{emb}}$ & Embedding dimension & \text{7168} \\
\hline
$n_\mathrm{hd}$ & Number of heads & \text{128} \\
\hline
$d_\mathrm{hd}$ & Head dimension & \text{128} \\
\hline
$d_\mathrm{dec}$ & Decompressed Q/K/V dimensions, $n_\mathrm{hd} \cdot d_\mathrm{hd}$ & \text{16384} \\
\hline
$d_\mathrm{Qco}, d_\mathrm{KVco}$ & Compressed Q, KV dimensions & \text{1536, 512} \\
\hline
$d_\mathrm{RoPE}$ & Rotary Positional Embedding dimension & \text{64} \\
\hline
$d_\mathrm{FFN}$ & FFN intermediate dimension & \text{18432} \\
\hline
$d_\mathrm{MoE}$ & MoE intermediate dimension & \text{2048} \\
\hline
$n_\mathrm{e}$ & Number of routed experts & \text{256} \\
\hline
$n_\mathrm{k}$ & Number of routed experts per token & \text{8} \\
\bottomrule
\end{tabular}
\vspace{-0.08in}
\end{table}

We analyze the computational characteristics of an MLA block, which comprises multiple FC layers and a core-attention layer, using the symbolic definitions in Table~\ref{tab:deepseek_r1_symbol}.

\subsection{Introducing a latent space to attention}

MLA employs a low-rank joint compression for the attention block, primarily reducing KV\$ capacity requirements while also lowering projection weight size and the associated FC layer computations.
It first compresses a hidden state matrix, mapping it to a lower-dimensional latent space (\textcircled{\scriptsize{a}} in Figure~\ref{fig:mlaflow_topdown}) to form a compressed Q ($\mathbf{C}_\mathrm{Q}$) and a compressed KV ($\mathbf{C}_\mathrm{KV}$) through projection using $\mathbf{W}_\mathrm{CQ}$ and $\mathbf{W}_\mathrm{CKV}$ (Eq.~\ref{eq:mla_latent}).
The resulting $\mathbf{C}_\mathrm{Q}$ and $\mathbf{C}_\mathrm{KV}$ are then decompressed (\textcircled{\scriptsize{c}}, \textcircled{\scriptsize{d}}, and \textcircled{\scriptsize{e}}) through projections using the corresponding decompression weights $\mathbf{W}_{\text{DX}_i}$, where $\mathbf{X} \in \{\mathbf{Q}, \mathbf{K}, \mathbf{V}\}$, to reconstruct the higher-dimensional full \ti{Q}, \ti{K}, and \ti{V}.
These are then used to perform the same core-attention layer as in a standard MHA block (Eq.~\ref{eq:mla_score} and Eq.~\ref{eq:mla_context}).

\vspace{-0.2em}
\begin{equation}
\label{eq:mla_latent}
\Scale[0.9]{
\underbrace{\mathbf{C}_{\mathrm{Q}}}_{\mathbb{R}^{\ell \times d_{\mathrm{Qco}}}}=\mathbf{H}_\ell \cdot \underbrace{\mathbf{W}_{\mathrm{CQ}}}_{\mathbb{R}^{d_{\mathrm{emb}}\times d_{\mathrm{Qco}}}}
,
\underbrace{\mathbf{C}_{\mathrm{KV}}}_{\mathbb{R}^{L \times d_{\mathrm{KVco}}}} = \mathbf{H}_L \cdot \underbrace{\mathbf{W}_{\mathrm{CKV}}}_{\mathbb{R}^{d_{\mathrm{emb}}\times d_{\mathrm{KVco}}}}
}
\end{equation}

\vspace{-0.2em}
\begin{equation}
\label{eq:mla_score}
\Scale[0.9]{
\begin{aligned} \mathbf{S}_{i} =  \mathbf{Q}_i \cdot (\mathbf{K}_{i})^{\text{T}} 
\! = \!(\mathbf{C}_\mathrm{Q} \cdot \underbrace{\mathbf{W}_{\mathrm{DQ}_i})}_{\mathclap{\mathbb{R}^{d_{\mathrm{Qco}} \times \frac{d_{\mathrm{dec}}}{n_{\mathrm{hd}}}}}} \cdot (\mathbf{C}_{\mathrm{KV}} \cdot \underbrace{ \mathbf{W}_{\mathrm{DK}_{i}}}_{\mathclap{\mathbb{R}^{d_{\mathrm{KVco}} \times \frac{d_{\mathrm{dec}}}{n_{\mathrm{hd}}}}}})^\text{T} 
& \end{aligned}
}
\end{equation}

\vspace{-0.4em}
\begin{equation}
\label{eq:mla_context}
\Scale[0.9]{
\mathbf{O}_{i} = \mathrm{Softmax}({\frac{\mathbf{S}_{i}}{\sqrt{d_{\mathrm{dec}}/{n_{\mathrm{hd}}}}}})\cdot (\mathbf{C}_{\mathrm{KV}} \cdot \underbrace{\mathbf{W}_{\mathrm{DV}_{i}}}_{\mathclap{\mathbb{R}^{d_{\mathrm{KVco}}\times \frac{d_{\mathrm{dec}}}{n_{\mathrm{hd}}}}}})
}
\end{equation}

\vspace{-0.2em}

\begin{table*}[t]
\centering
\renewcommand{\arraystretch}{1.3}
\setlength{\tabcolsep}{2pt}
\caption{Comparison of MLA with and without the reordering in terms of FLOPS, memory access, and \ai for prefill and decode stages assuming $B, L \gg n_\mathrm{hd}, d_\mathrm{hd}$.}
\vspace{-0.1in}
\label{tab:reorder_flops_memaccess}
\resizebox{0.94\textwidth}{!}{%
\begin{tabular}{|c|c|c|l|c|c|c|}
\hline
\textbf{Layer} & \textbf{Phase} & \textbf{Reordering} & \multicolumn{1}{c|}{\textbf{FLOPs}} & \textbf{Asymptotic Memory Access} & \textbf{\ai} & \textbf{\ai in DeepSeek-R1} \\
\hline

\multirow{4}{*}{Prefill}
& \multirow{2}{*}{K decompress}
& without 
& $B2L d_{\mathrm{KVco}} n_\mathrm{hd}d_\mathrm{hd}$ 
& $2BLn_\mathrm{hd}d_\mathrm{hd}$ 
& $\approx d_{\mathrm{KVco}}$ & $\approx 512$ \\
\cline{3-7}
& 
& with
& $B2L d_{\mathrm{KVco}} n_{\mathrm{hd}}d_\mathrm{hd}$ 
& $2B(L n_{\mathrm{hd}}d_{\mathrm{hd}} + Ln_{\mathrm{hd}}d_{\mathrm{KVco}})$ 
& $\approx (d_\mathrm{hd}^{-1}+d_\mathrm{KVco}^{-1})^{-1}$ & $\approx 100$\\
\cline{2-7}
& \multirow{2}{*}{Score}
& without  
& $B2n_{\mathrm{hd}}L^2 d_{\mathrm{hd}}$ 
& $2Bn_{\mathrm{hd}}L^2$ 
& $\approx d_{\mathrm{hd}}$ & $\approx 128$\\
\cline{3-7}
&
& with
& $B2n_\mathrm{hd}L^2 d_{\mathrm{KVco}}$ 
& $2Bn_{\mathrm{hd}} L^2$ 
& $\approx d_{\mathrm{KVco}}$ & $\approx 512$\\
\cline{2-7}
\hline
\multirow{4}{*}{\makecell{Decode}}
& \multirow{2}{*}{K decompress}
& without 
& $B2d_{\mathrm{KVco}}L n_\mathrm{hd}d_\mathrm{hd}$ 
& $2BLd_{\mathrm{dec}}$ 
& $\approx d_{\mathrm{KVco}}$ & $\approx 512$\\
\cline{3-7}
& 
& with
& $B2d_\mathrm{KVco}n_\mathrm{hd}d_\mathrm{hd}$ 
& $2B d_{\mathrm{KVco}}n_\mathrm{hd}$ 
& $\approx d_{\mathrm{hd}}$ & $\approx 128$\\
\cline{2-7}
& \multirow{2}{*}{Score}
& without  
& $B2n_{\mathrm{hd}}L d_{\mathrm{hd}}$ 
& $2Bn_{\mathrm{hd}}d_{\mathrm{hd}}L$ 
& $\approx 1$ & $\approx 1$ \\
\cline{3-7}
& 
& with
& $B2n_{\mathrm{hd}}L d_{\mathrm{KVco}}$ 
& $2B(d_{\mathrm{KVco}} L + n_{\mathrm{hd}}L)$ 
& $\approx (n_\mathrm{hd}^{-1}+d_\mathrm{KVco}^{-1})^{-1}$ & $\approx 100$\\
\hline

\end{tabular}
\vspace{-0.2in}
}
\end{table*}

The attention block's parameter footprint shrinks substantially both in absolute size and as a fraction of total model parameters.
Employing the low-rank joint compression reduces the weight used to generate $\mathbf{K}$ and $\mathbf{V}$ from $d_\mathrm{emb} \times d_\mathrm{dec} = 7K \times16K=112M$ to $d_\mathrm{emb} \times d_\mathrm{KVco} + d_\mathrm{KVco} \times d_\mathrm{dec} = 7k \times 0.5K + 0.5K\times16K=11.5M$ (see Table~\ref{tab:deepseek_r1_symbol} for notations). 
\emph{With this reduced weight burden, replicating its FC layer parameters across devices is far more tractable, making DP for the attention block a compelling choice.}

For conventional LLMs, a key limiting factor of the batch size is KV\$ size. 
MLA introduces a latent space on attention, drastically reducing the KV\$ ($\mathbf{C}_\mathrm{KV}$) storage (Figure~\ref{fig:GPT_DeepSeek_mem_usage}).
For conventional LLMs, $d_\mathrm{dec}(=d_\mathrm{emb})$-dimensioned K and V are cached per layer. 
In DeepSeek-R1, K and V share the same compressed cache with a dimension of $d_\mathrm{KVco} + d_\mathrm{RoPE}$, which is usually an order of magnitude smaller than both $d_\mathrm{emb}$ and $d_\mathrm{dec}$.
In GPT-3, KV\$ consumes 4.5MB ($= d_{\mathrm{dec}} \times n_{\mathrm{decoder}} \times \text{(K \& V)} \times \text{FP16} = 12288\! \times\! 96 \! \times \! 2 \! \times \! 2\text{B}$) per token, whereas $\mathbf{C}_\mathrm{KV}$ in DeepSeek-R1 requires only 68.6KB ($=\!(d_{\mathrm{KVco}}\!+\!d_{\mathrm{RoPE}}) \times n_{\mathrm{decoder}} \times \text{BF16} = 576\! \times\! 61\! \times\! 2\text{B}$) per token.

Smaller KV\$ size allows larger batch sizes for FC layers, which improve compute utilization.
However, MLA shares the same core-attention layer structure as conventional MHA, suffering from extremely low \ai ($\approx \!1$).
Also, $\mathbf{C}_\mathrm{KV}$ decompression needs to be computed on demand during runtime.

Nevertheless, MLA enables \emph{layer reordering} (or simply \emph{reordering})~\cite{arxiv-2024-deepseek-v2} to improve data reuse by rearranging the layers in the attention block.
One of MLA's key features is decoupled RoPE: 
Instead of applying RoPE directly to Q and K, RoPE is computed separately, the result of which is added element-wise in the score layer.
By removing the nonlinearity between the QKV generation layer and the core-attention layer, Eq.~\ref{eq:mla_score} can be algebraically rewritten:

\begin{equation}
\label{eq:mla_reorder_score1}
\Scale[1.0]{
\begin{split}
\mathbf{S}_i &= \mathbf{Q}_{i} \cdot ( \mathbf{C}_{\mathrm{KV}} \cdot  \mathbf{W}_{\mathrm{DK}_{i}})^\text{T} \\
&= \mathbf{Q}_i \cdot ({\mathbf{W}_{\mathrm{DK}_i}}^{\mathrm{T}} \cdot {\mathbf{C}_{\mathrm{KV}}}^{\text{T}}) \\
&= (\mathbf{Q}_i \cdot {\mathbf{W}_{\mathrm{DK}_i}}^{\text{T}}) \cdot {\mathbf{C}_{\mathrm{KV}}}^{\text{T}} \\
\mathbf{S} &= 
\begin{bmatrix} 
\mathbf{Q}_1 \cdot {\mathbf{W}_{\mathrm{DK}_1}}^{\text{T}} \\
\mathbf{Q}_2 \cdot {\mathbf{W}_{\mathrm{DK}_2}}^{\text{T}} \\
\vdots \\
\mathbf{Q}_{n_{\mathrm{hd}}} \cdot {\mathbf{W}_{\mathrm{DK}_{n_{\mathrm{hd}}}}}^{\mathbf{T}}
\end{bmatrix} \cdot {\mathbf{C}_{\mathrm{KV}}}^{\text{T}}
\end{split}}
\end{equation}

\noindent
A similar reordering can also be applied to the context layer:
\begin{equation}
\label{eq:mla_reorder_context1}
\Scale[1.0]{
\begin{split}
\mathbf{O}_{i} &= \text{Softmax}({\frac{\mathbf{S}_{i}}{\sqrt{d_{\mathrm{dec}}/{n_{\mathrm{hd}}}}}}) \cdot (\mathbf{C}_{\mathrm{KV}} \cdot \mathbf{W}_{\mathrm{DV}_i}) \\
&= \! (\text{Softmax}({\frac{\mathbf{S}_{i}}{\sqrt{d_{\mathrm{dec}}/{n_{\mathrm{hd}}}}}}) \cdot \mathbf{C}_{\mathrm{KV}}) \cdot \mathbf{W}_{\mathrm{DV}_i} 
\end{split}}
\end{equation}

\begin{figure*}[!ht]
\centering
\includegraphics[width=0.99\textwidth]{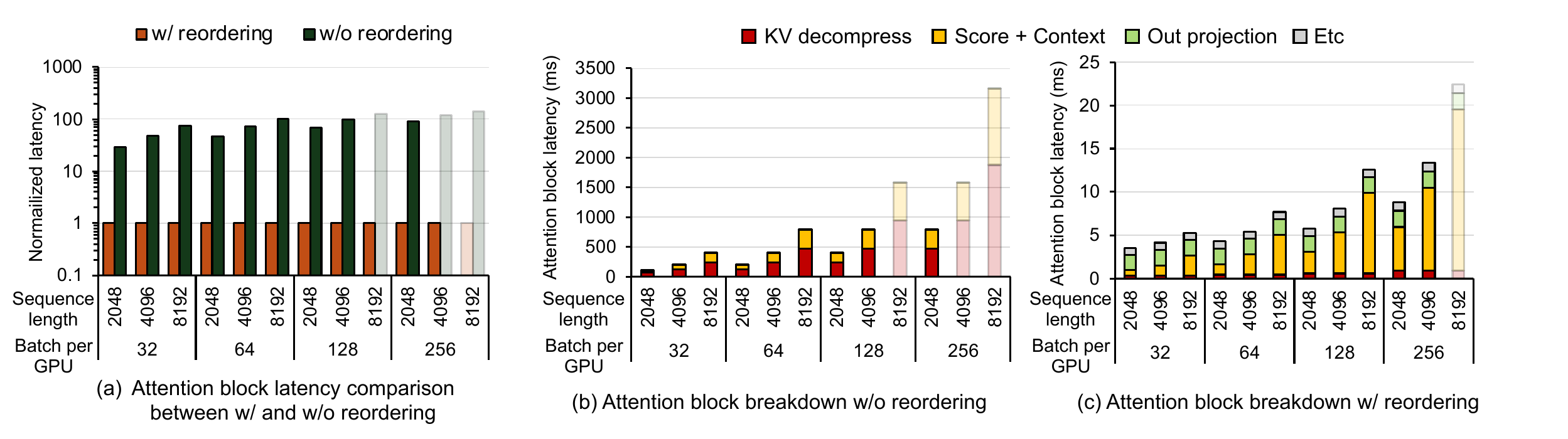}
\vspace{-0.0in}
  \caption{(a) Normalized latency of the attention block in the decode stage without reordering compared to a reordered attention block. (b) and (c) show the execution time ratio of each layer in the attention block in the decode stage with and without reordering, across varying sequence length and batch size. All experiments assumed a 32 B200 GPU system. See \S\ref{sec:experimental_setup} for our experimental setup in detail.}
  \label{fig:attention_time_breakdown}
  \vspace{-0.0in}
\end{figure*}

\subsection{Impact of reordering MLA}
\label{impact_of_reordering}

The \emph{reordering}\footnote{Although the DeepSeek papers~\cite{arxiv-2024-deepseek-v2, arxiv-2024-deepseek-v3} refer to this technique as ``absorption,'' we use this term to distinguish it from weight fusion, which merges multiple weight matrix multiplications into a single computational step.} optimization improves hardware utilization and drastically reduces the latency of attention during the decode stage.
However, it rather increases the latency during the prefill stage.
This is because reordering in MLA significantly changes the computational characteristics, such as FLOPs, the number of memory accesses, and \ai.

Table~\ref{tab:reorder_flops_memaccess} compares the FLOPs and memory requirements in K decompression and score layers across both prefill and decode stages when a single \dev is used, with and without reordering.
V decompression and context layers exhibit similar trends regardless of reordering because $\mathbf{K}$ and $\mathbf{W}_\mathrm{DK}$ share the same structure with $\mathbf{V}$ and $\mathbf{W}_\mathrm{DV}$, respectively.
This analysis leads to a number of notable observations.

Without reordering, the core-attention layer in MLA preserves the same computational flow as MHA, except for the runtime compression and decompression layers, which generate Q, K, and V.
The required amounts of computations (FLOPs) and memory accesses for the K decompression and score layers are proportional to $B$ and $L$ during the decode stage (Table~\ref{tab:reorder_flops_memaccess}), whereas those of the other FC layers do not scale with $L$.
As a result, KV decompression and core-attention layers dominate the execution time of an attention block as $L$ increases (Figure~\ref{fig:attention_time_breakdown}(b)). 
At $B\! =\! 128$ and $L\! =\! 4096$, KV decompression and core-attention layers account for 59\% and 40\% of the attention block latency, respectively.

\vspace{-0.00in}
\begin{tcolorbox}[boxsep=0pt,left=3pt,right=3pt,arc=0pt]
\emph{\textbf{Observation-1:}}
In the decode stages of MLA without reordering, KV decompression and core-attention layers dominate the runtime of MLA's attention blocks.
\end{tcolorbox}
\vspace{-0.00in}

After applying layer reordering, MLA multiplies $\mathbf{W}_\mathrm{DK}$ with $\mathbf{Q}$.
Because the size of $\mathbf{Q}$ is independent of $L$ in the decode stage, layer reordering eliminates the need to decompress the entire $\mathbf{C}_\mathrm{KV}$, reducing the cost of $\mathbf{K}$ decompression by a factor of $L$ (Table~\ref{tab:reorder_flops_memaccess}).
The portion of K decompression---previously the dominant component---has been significantly reduced due to reordering (see Figure~\ref{fig:attention_time_breakdown}(c)).
By contrast, this benefit disappears in the prefill stage as the size of $\mathbf{Q}$ is proportional to $L\! =\! L_\mathrm{in}$, leaving the computational cost of K decompression unchanged.

Layer reordering increases computations required for the score layer, which is a part of the core-attention layer, by $\sfrac{d_\mathrm{KVco}}{d_\mathrm{hd}}$ (4 for DeepSeek-R1) times in both prefill and decode stages.
The original score layer is replaced by a multiplication between $\mathbf{Q}_i \cdot {\mathbf{W}_\mathrm{DK_i}}^\mathrm{T}$ and ${\mathbf{C}_\mathrm{KV}}^\mathrm{T}$ (Eq.~\ref{eq:mla_reorder_score1}), where one of the matrix dimensions becomes $d_\mathrm{KVco}$ instead of $d_\mathrm{hd}$.

\vspace{-0.00in}
\begin{tcolorbox}[boxsep=0pt,left=3pt,right=3pt,arc=0pt]
\emph{\textbf{Observation-2:}}
While layer reordering maintains or reduces the FLOPs of KV decompression, it increases the FLOPs of core-attention layers in both prefill and decode stages.
\end{tcolorbox}
\vspace{-0.00in}

With layer reordering in the decode stage, the core-attention layer reads $\mathbf{C}_\mathrm{KV}$ instead of decompressed KV\$, reducing memory access by $d_\mathrm{dec}/d_\mathrm{KVco}$.
Because $\mathbf{C}_\mathrm{KV}$ can be shared among the heads,
the \ai of both score and context layers reaches approximately $n_\mathrm{hd}d_\mathrm{KVco}/(n_\mathrm{hd} + d_\mathrm{KVco})$ ($\approx$ 100 for DeepSeek-R1) in the decode stages.

FlashMLA~\cite{github-2025-flashmla}, a GPU-optimized implementation, further doubles this Op/B by reusing ${\mathbf{C}_\mathrm{KV}}$ loaded during the score layer in the subsequent context layer.
The resulting doubled \ai 
(\eg, $\approx$ 200 Op/B in DeepSeek-R1) closely approaches the ridge point $\mathrm{RP_{acc}}$ of modern \devs, exhibiting a \emph{balance} between computation and memory bandwidth for modern \devs.

Despite using a latent space, explicitly generating decompressed $\mathbf{K}$ and $\mathbf{V}$ requires a significant amount of memory for activation, limiting $B$ as shown in Figure~\ref{fig:attention_time_breakdown}.
For DeepSeek-R1, decompressing the $\mathbf{K}$ tensor with a per-accelerator batch size of 256 and $L=4096$ inflates the activation footprint to $\approx$50GB.
As layer reordering shrinks the size of this on-the-fly activation, the maximum feasible $B$ increases.
The increased $B$ delivers proportionally higher throughput on FC layers.

\vspace{-0.00in}
\begin{tcolorbox}[boxsep=0pt,left=3pt,right=3pt,arc=0pt]
\emph{\textbf{Observation-3:}}
Layer reordering improves hardware utilization by increasing the \ai of the core-attention layer to approach $\mathrm{RP_{acc}}$ of modern accelerators in the decode stage and by enabling sufficient batching for FC layers.
\end{tcolorbox}
\vspace{-0.00in}

As the \ai of the core-attention layer, which dominates the runtime of attention blocks, approaches $\mathrm{RP_{acc}}$,
the time spent on computations remains approximately equal to the time spent on accessing $\mathbf{C}_\mathrm{KV}$.
Thus, the latency of the core-attention layer is reduced approximately by $2d_\mathrm{dec}/d_\mathrm{KVco}$ ($=\! 64$ for DeepSeek-R1) after layer reordering.

In contrast, reordering increases the latency of attention blocks during the prefill stage.
Layer reordering expands the activation's dimension per head from $d_\mathrm{hd}$ to $d_\mathrm{KVco}$. 
For the KV decompression layer, while the FLOPs remains unchanged, the number of memory accesses increases, leading to longer execution times.
Also, the FLOPs increases for the core-attention layer (\emph{Obs.2}).

Putting it all together, layer reordering significantly reduces the latency of attention blocks in the decode stage by up to 103.12$\times$ (see Figure~\ref{fig:attention_time_breakdown}).
By contract, the attention block latency in the prefill stage slows down by up to $2.21\times$ with layer reordering.
Hereafter, the prefill stage uses MLA without reordering and the decode stage uses MLA with reordering.

\vspace{-0.00in}
\begin{tcolorbox}[boxsep=0pt,left=3pt,right=3pt,arc=0pt]
\emph{\textbf{Observation-4:}}
Layer reordering substantially reduces attention block latency by reducing memory access and drastically mitigating the impact of KV decompression overhead.

\end{tcolorbox}
\vspace{-0.00in}

\begin{figure}[!tb]
\centering
\includegraphics[width=0.99\columnwidth]{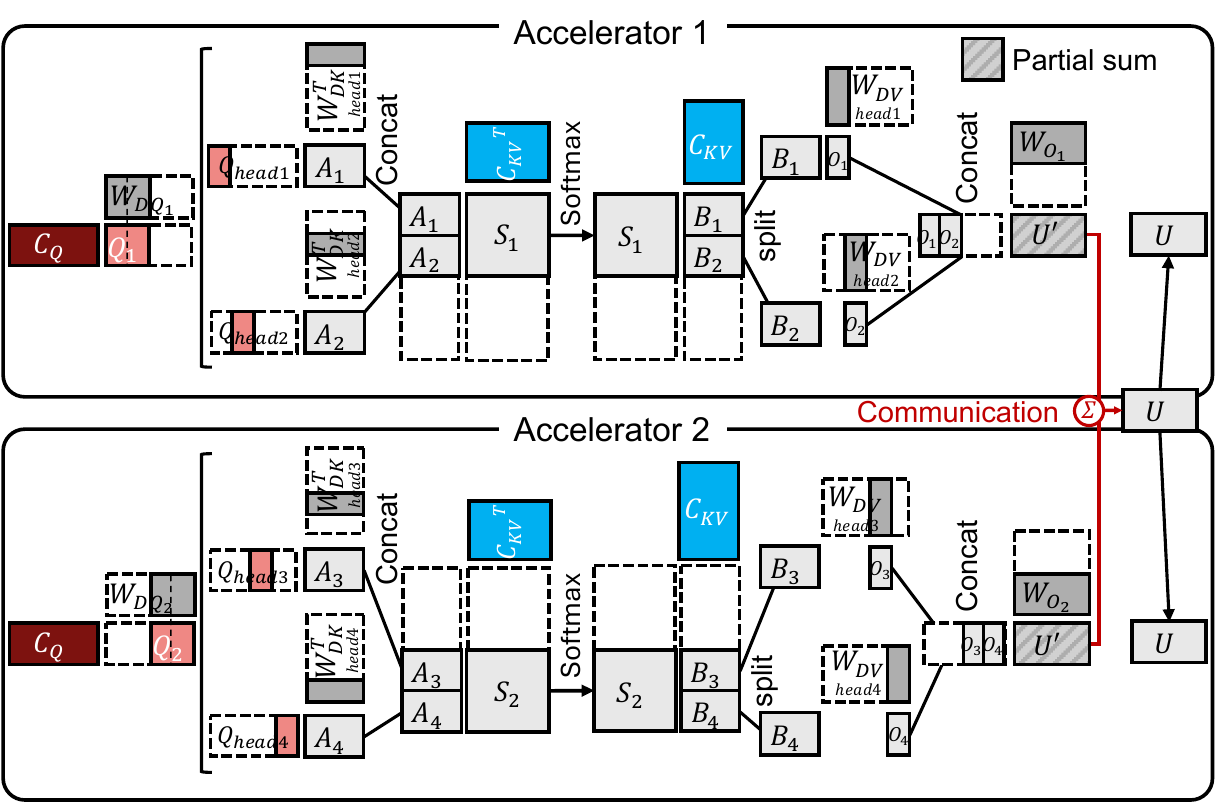}
  \vspace{-0.07in}
  \caption{Computation flow of an MLA attention block with ${deg}_\mathrm{TP}=2$ and $n_\mathrm{hd}=4$ on our experimental setup.}
  \label{fig:mla_parallelism}
  \vspace{-0.07in}
\end{figure}

\begin{figure}[!bt]
\centering
\includegraphics[width=0.99\columnwidth]{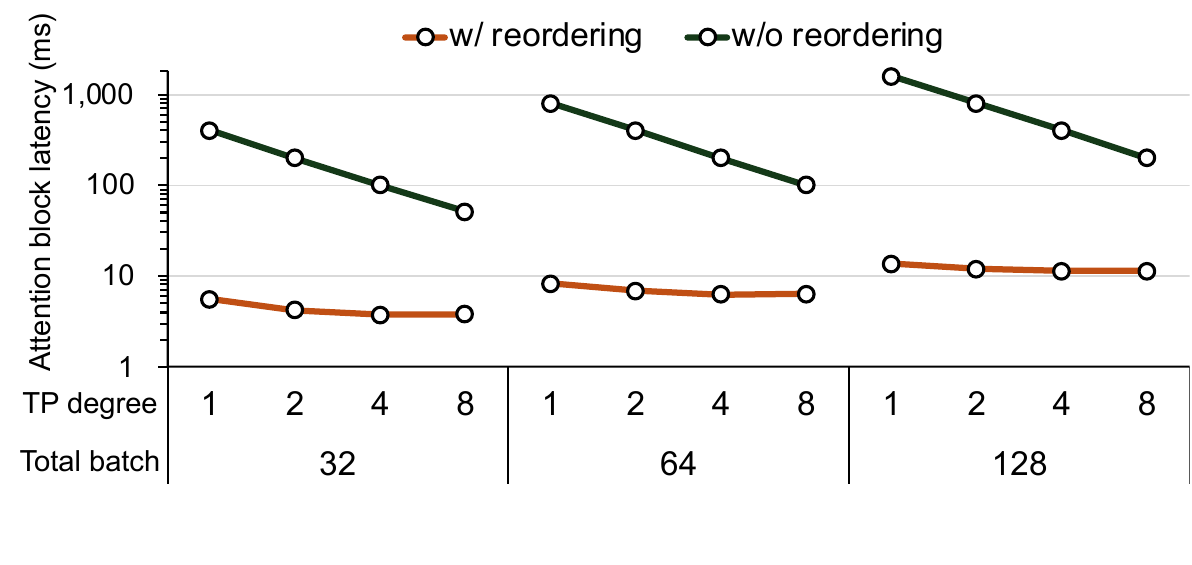}
\vspace{-0.0in}
  \caption{
    Latency of the attention block with and without reordering as batch sizes and $\mathrm{deg}_\mathrm{TP}$ vary when $L\! =\! 4096$.}
  \label{fig:mla_latency_reorder_vs_base_tp}
\vspace{-0.1in}
\end{figure}

\subsection{Parallelism on MLA}
\label{sec:MLA_parallelism}
TP offers little latency benefit once layer reordering is applied to the attention block.
Although reordering drastically lowers the core-attention layer latency, the latency still scales with $B$ and $L$ in the decode stage and becomes dominant at large $B$ and $L$ values (Figure~\ref{fig:attention_time_breakdown}(c)).
When heads are independent, as in MHA, head-wise TP can distribute the KV\$ across multiple \devs, reducing latency.
Also, the \ai is preserved within a head as it is not affected by TP.

In the reordered MLA, however, all heads share the same $\mathbf{C}_\mathrm{KV}$; thus, all \devs need to store and read the whole $\mathbf{C}_\mathrm{KV}$, nullifying the performance and capacity gains from scaling out to \tut{deg}{TP} \devs.
Also, TP reduces the number of heads per \dev, thereby reducing the \ai by \tut{deg}{TP}.
As depicted in Figure~\ref{fig:mla_parallelism}, TP reduces the number of heads batched on each \dev, thereby reducing the \ai by \tut{deg}{TP}.

Figure~\ref{fig:mla_latency_reorder_vs_base_tp} compares the latency impact of TP on reordered and non-reordered MLA attention blocks. 
Although the FC layers in an attention block benefit from TP, the core-attention layer dominates the runtime and limits improvements in the non-reordered case. 
Thus, using DP alone is preferable to combining TP and DP in the attention block.

\vspace{-0.0in}
\begin{tcolorbox}[boxsep=0pt,left=3pt,right=3pt,arc=0pt]
\emph{\textbf{Observation-5:}}
In the decode stage with reordering, tensor parallelism fails to provide a meaningful latency reduction.
\end{tcolorbox}
\vspace{-0.1in}

\section{Insights on Mixture of Experts}
\label{sec:contribution-3}

\subsection{Mixture of Experts (MoE)}
Although it is a common belief that larger models with more parameters produce higher-quality responses~\cite{arxiv-2020-scalinglaw}, the substantial computational overhead associated with scaling LLMs hinders further growth. 
MoE~\cite{jmlr-2022-switchtransformer} addresses this issue by employing sparse activation in FFN blocks;
MoE introduces a pool of \emph{experts} and activates only a small subset of experts for each input.
Recent LLMs~\cite{arxiv-2024-deepseek-v3, arxiv-2025-deepseek-r1, meta-2025-llama4} adopt a hybrid architecture consisting of two types of experts: a \emph{shared expert} and \emph{routed experts}. 
The former is activated for every token during inference, whereas the latter are selectively activated based on a routing mechanism that dynamically assigns $n_\mathrm{k}$ experts out of $n_\mathrm{e}$ experts to each token.
The computational procedure of an MoE block can be described as
\begin{equation}
\label{eq:moe procedure}
\Scale[1.0]{
\begin{split}
\text{MoE}(\mathbf{u})\!=\! \Big(\sum_{e \in \{1,\cdots, n_e\}} \text{Expert}_{e}(\mathbf{u})\Big)\! +\! \text{Expert}_\mathrm{shared}(\mathbf{u})
\end{split}
}
\end{equation}
\noindent By utilizing only $n_k$ experts ($n_k < n_e$), along with one shared expert per token at runtime, MoE effectively scales the model with low computational overhead.
$n_k$ and $n_e$ vary across  models: 
DeepSeek-R1~\cite{arxiv-2025-deepseek-r1} employs eight routed experts selected from a pool of 256 while Llama4-Maverick~\cite{meta-2025-llama4} uses a single routed expert selected from a pool of 128.


The execution time of an MoE block is dominated by expert computations and by the communication required to dispatch/combine tokens to/from the selected experts~\cite{mlsys-2025-comet, neurips-2024-moe-efficient}.
A common strategy to scale MoE inference is exploiting expert parallelism (EP), which distributes experts across \devs. 
Under EP, tokens are transferred to the devices holding their routed experts; after the computation, the results are combined with additional communication overhead.
As different types of parallelism can be selectively applied at the block level, we use $deg_\mathrm{EP}$ to denote the degrees of EP.

Moreover, 
Using EP for MoE in a multi-\dev system introduces complications regarding batching, where we need to maximize the usage of each \dev's arithmetic throughput
while also satisfying the memory capacity limitations and SLO constraints.
Meanwhile, communication overhead between the \devs in a serving system is highly dependent on the system-wide interconnect (\eg, NVLink~\cite{ieeemicro-2017-pascalnvlink,nvidia-b200}, InfiniBand~\cite{infiniband-whitepaper,nvidia-infiniband-x800}, and optical links~\cite{sigcomm-2022-jupiter}) specification. 
In this section, we analyze the impact of both factors on the performance of MoE blocks.

\subsection{Maximize compute utilization in MoE blocks}
\label{sec:contribution_expert}
In a multi-\dev system, an efficient expert computation requires careful design to make the best use of each \dev's arithmetic throughput by batching tokens for each expert.
Although attention and MoE blocks both contain FC layers, their effective $B$'s differ due to the MoE sparsity and distinct parallelization strategies.
For attention, requests are split across $deg_\mathrm{DP}$ DP groups and each group processes $B/deg_\mathrm{DP}$ requests using TP; the \ai of the FC layers in an attention block scales with $B/deg_\mathrm{DP}$.

In contrast, as we utilize EP for MoE, each expert handles $B \cdot n_\mathrm{k}/n_\mathrm{e}$ tokens on average.
Because tokens are dynamically routed to experts at runtime, the number of tokens assigned to each expert can vary across experts.
Hereafter, we denote by $\Gamma_{imb}$ \emph{the load imbalance ratio}, defined as the ratio between the actual number of tokens processed by an expert and the number for an ideal uniform distribution.
Considering the load imbalance, the \ai of each expert would be $\Gamma_{imb} \cdot B \cdot n_\mathrm{k}/n_\mathrm{e}$.

For analytical simplicity, we conduct the following analysis based on the average behavior of all experts. 
The effects of load imbalance will be deeply discussed in \S\ref{sec:eval:expert_skew}.
To reach the ridge point $\mathrm{RP_{acc}}$ for FC layers in each block, the batch size $B$ must satisfy:

\begin{equation}
\label{eq:moe_max_batchsize1}
\Scale[0.95]{
\begin{split}
\mathrm{B} \ge \mathrm{B_{attn}} &= \mathrm{RP_{acc}} \cdot \mathrm{deg_{DP}} \\
\mathrm{B} \ge \mathrm{B_{MoE}} &= \mathrm{RP_{acc}} \cdot \frac{n_e}{n_k}
\end{split}}
\end{equation}

\noindent $\mathrm{B_{attn}}$ and $\mathrm{B_{MoE}}$ are the batch sizes that reaches $\mathrm{RP_{acc}}$ for the FC layers of an attention and an MoE block, respectively.
We denote the minimum $B$ that satisfies Eq.~\ref{eq:moe_max_batchsize1} as $\mathrm{B_{RP}}=\text{max}(\mathrm{B_{attn}},\mathrm{B_{MoE}})$.

While $\mathrm{B_{attn}}$ is influenced by $\mathrm{deg_{DP}}$, $\mathrm{B_{MoE}}$ only depends on the model and the target \dev and is independent of the number of accelerators.
Since $n_e$ and $n_k$ are model parameters, the \ai of the FC layers in an MoE block is determined once $\mathrm{RP_{acc}}$ and the model are fixed.

\vspace{-0.0in}
\begin{tcolorbox}[boxsep=0pt,left=3pt,right=3pt,arc=0pt]
\emph{\textbf{Observation-6:}}
$B_{\mathrm{MoE}}$ is determined once the model and the target accelerator are fixed.
\end{tcolorbox}
\vspace{-0.1in}

\subsection{Two primary factors limiting batch size}
\label{sec:batch_size_limiting_factor}

\noindent While batching $\mathrm{B_{RP}}$ requests is desirable, the feasible batch size is limited by two factors: memory capacity and SLO.

\noindent \textbf{Memory capacity}: To fully utilize the \dev's computational resources, data must be served at high bandwidth.
To achieve that, the entire working set should reside in the main memory (\eg, HBM).
This working set includes the weights for attention and MoE blocks, as well as KV\$.
As the weight size is predetermined, serving systems typically use the remaining memory for activation and KV\$, whose sizes are proportional to $B$.
Thus, the memory space requirements for model weights determine the maximum feasible batch size ($\mathrm{B_{cap}}$) as follows:
\begin{equation}
\label{eq:moe_max_batchsize_cap}
\Scale[0.95]{
\begin{split}
\mathrm{B_{cap}}
&= \frac{\mathrm{M_{cap}} \cdot n_\mathrm{acc} - n_{decoder} \cdot (\mathrm{M_{attn}}\cdot{\mathrm{deg_{DP}}} + \mathrm{M_{MoE}})}{{n_{decoder} \cdot \mathrm{M_{KV}} \cdot L} + \mathrm{M_{act}}(L)}
\end{split}}
\end{equation}

\noindent where $\mathrm{M_{cap}}\cdot n_\mathrm{acc} $ denotes the memory capacity of a system composed of $n_{acc}$ \devs, each having a $\mathrm{M_{cap}}$ capacity. 
$\mathrm{M_{attn}}$ and $\mathrm{M_{MoE}}$ represent the model weight sizes of a single decoder block's attention and MoE, respectively.
We denote the KV\$ size per token for each decoder block as $\mathrm{M_{KV}}$.
The activation memory space required by a decoder block per token on each \dev, $\mathrm{M_{act}}$, depends on the sequence length $L$.
As this memory space is reused across multiple decoder blocks, the $\mathrm{M_{act}}(L)$ term in Eq.~\ref{eq:moe_max_batchsize_cap}
does not scale with $n_{\mathrm{decoder}}$.
To batch $\mathrm{B_\mathrm{RP}}$ requests (from the previous section), $\mathrm{B_{cap}}$ should be greater than $B_\mathrm{RP}$.

\noindent \textbf{SLO}: As excessive batching would incur latency overheads, SLO becomes another limiting factor for feasible batch sizes. 
In a disaggregated system, the time per output token (TPOT), a key latency metric in LLM serving, is determined by the latency of each decode stage and expressed as follows:
\begin{equation}
\label{eq:tpot}
\Scale[0.9]{
\begin{split}
\mathrm{TPOT(B,L)}
\!&=\!n_{\mathrm{decoder}} \cdot \left(\underbrace{\frac{\mathrm{M_{attn}} \cdot {\mathrm{deg_{DP}}} + \mathrm{M_{MoE}} }{n_{\mathrm{acc}} \cdot \mathrm{BW_{Mem}}}}_{\text{model load lat.}}\!+\!\underbrace{\mathrm{\delta(B, L)}}_{\text{additional lat.}} \right)\notag
\end{split}}
\end{equation}

\noindent where both the first and the second terms in the parentheses represent latencies for each decoder block: the first accounts for the latency to read model weights and the second, $\mathrm{\delta(B,L)}$, includes additional latency such as memory access time for the KV\$ and activations, communication overhead, and any remaining computation time.
The additional latency term is a function of $B$ and $L$. 

As the memory access time for the KV\$ and activations, along with communication time, is unavoidable when processing each decoder block, the minimum bound of this additional latency, $\mathrm{\delta_{min}(B, L)}$, is given by
\begin{equation}
\label{eq:tpot_additional_latency}
\Scale[0.9]{
\begin{split}
\mathrm{\delta_{min}(B,L)}
&\ge\mathrm{B} \cdot \left( \frac{\mathrm{M_{KV}} \cdot \mathrm{L} + \mathrm{M_{act}}(L)}{n_{\mathrm{acc}} \cdot \mathrm{BW_{mem}}}\right) + \mathrm{Comm(B,L)}
\end{split}}
\end{equation}

\begin{table}[!tb]
\caption{Model configuration used in evaluation. Both Llama4-Maverick and DeepSeek-R1 have 1 shared expert.}
\label{tab:model_configuration_deployment}
\vspace{-0.1in}
\centering
\resizebox{\columnwidth}{!}{
\begin{tabular} {l|l|l|l|l|l|l|l|l|l|l}
\toprule
\textbf{Model} & \# of par. & \textbf{$d_{\mathrm{emb}}$} & 
\textbf{$deg_{\mathrm{grp}}$} & \textbf{$d_{\mathrm{FFN}}$} & \textbf{$d_{\mathrm{MoE}}$} & \textbf{$n_{\mathrm{e}}$} & \textbf{$n_{\mathrm{k}}$} & \textbf{$deg_{\mathrm{TP}}$} & \textbf{$deg_{\mathrm{DP}}$} & \textbf{$deg_{\mathrm{EP}}$}\\
\midrule
GPT-3 & 175B & \text{12K} &  \text{1} & \text{48K} & \text{-} & \text{-} & \text{-} & \text{8} & \text{4} & \text{-}\\
\hline
Llama4 & 400B &  \text{5K} & \text{5}  & \text{16K} & \text{8K} & \text{1} & \text{128} & \text{8} & \text{4} & \text{32}\\
\hline
DeepSeek-R1& 671B & \text{7K} & \text{1} & \text{18K} & \text{2K} & \text{8} & \text{256} & \text{1} & \text{32} & \text{32} \\
\bottomrule
\end{tabular}
}
\vspace{-0.08in}
\end{table}

\noindent where $\mathrm{Comm(B, L)}$ denotes the communication overhead between the \devs.
Increasing $B$ leads to larger KV\$ and activation sizes, thus increasing the minimum bound of TPOT.
The theoretical maximum batch size, $\mathrm{B_{SLO}}$, that satisfies the SLO time limit ($\mathrm{TPOT_{SLO}}$) can be achieved under the minimum latency.
A batch size B greater than $\mathrm{B_{SLO}}$ can never satisfy the $\mathrm{TPOT_{SLO}}$ time limit, establishing an upper bound on the feasible batch size.

\vspace{-0.00in}
\begin{tcolorbox}[boxsep=0pt,left=3pt,right=3pt,arc=0pt]
\emph{\textbf{Observation-7:}}
While MoE tightens the batch size limit due to weight overheads, MLA complements this through its small KV\$ size.
\end{tcolorbox}
\vspace{-0.00in}
MoE weights ($\mathrm{M_{MoE}}$) are typically larger than the FFN weights in standard LLMs, increasing memory requirements for the model weights.
Then, $\mathrm{B_{cap}}$ decreases as less memory space remains for KV\$ (see Eq.~\ref{eq:moe_max_batchsize_cap}).
It also increases the model load latency, which shortens the time available for $\mathrm{\delta_{min}(B_{SLO}, L)}$, thereby reducing $\mathrm{B_{SLO}}$. 
In contrast, the reduction of $\mathrm{M_{KV}}$ and $\mathrm{M_{attn}}$ by MLA enables storing the KV\$ for more requests in the main memory, thereby increasing $\mathrm{B_{cap}}$.
It also reduces the load time for $\mathrm{M_{KV}}$ and $\mathrm{M_{attn}}$, allowing higher $\mathrm{B_{SLO}}$ (see Eq.~\ref{eq:tpot_additional_latency}).
Thus, as for the batch size limits, MLA and MoE impose complementary effects. 

\begin{figure*}[!tb]
  \center
  \includegraphics[width=0.99\textwidth]{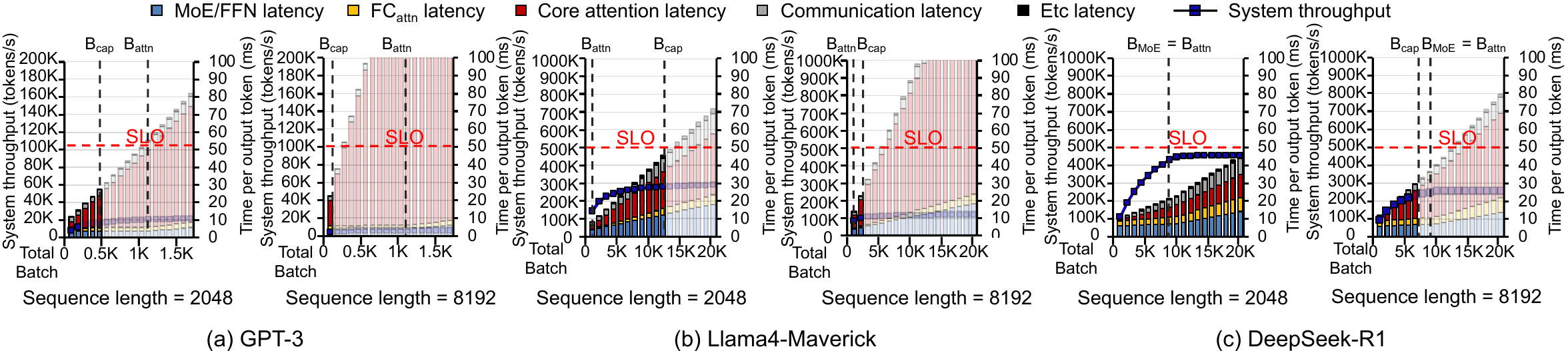}
  \vspace{-0.1in}
  \caption{Throughput-latency graph for the decode stages of GPT-3, Llama4-Maverick, and DeepSeek-R1. We assume a 32 B200 GPU system.}
  \label{fig:throughput_latency_32dev}
  \vspace{-0.1in}
\end{figure*}

\subsection{Communication cost}
\label{sec:communication_cost}
To reduce MoE execution time, both interconnect bandwidth and expert-distribution skew must be addressed.
Besides expert computations, communication is a dominant contributor to the overall MoE execution time~\cite{neurips-2024-moe-efficient}.
As experts are distributed across multiple accelerators using EP, the system must dispatch tokens to, and combine tokens from, the selected experts.
As tokens are transferred over the interconnect between the accelerators, communication time is determined by the size of the transferred tokens and the available interconnect bandwidth.
In the MoE blocks, the communication time varies across the \devs because each \dev sends or receives a different amount of data due to the imbalance in expert distributions.

\begin{equation}
\label{eq:moe_comm}
\Scale[0.85]{
\begin{split}
\mathrm{Comm_{MoE}(B)} &= 2 \cdot \max_{a \in Acc.} \left( \Gamma_{imb}^{acc}(a) \right) \cdot \frac{\mathrm{M_{token}} \cdot n_{k} \cdot {\mathrm{B}}}{\mathrm{BW_{Int}} \cdot n_{acc}} \!+\!\mathrm{\alpha}
\end{split}}
\end{equation}

Eq.~\ref{eq:moe_comm}\footnote{We assume a fully connected, switch-based interconnect topology that offers uniform bidirectional bandwidth ($\mathrm{BW_{Int}}$) among all accelerators. $\mathrm{M_{token}}$ represents the size of token by a single decoder.
$\alpha$ denotes the additional latency in the network.} provides a simplified model of the MoE communication time. 
$\Gamma_{imb}^{acc}$ denotes the load imbalance ratio at the accelerator level, computed from the total number of tokens processed by all the experts assigned to an accelerator.
When the batch size increases, the interconnect bandwidth and $\Gamma_{imb}^{acc}$ become the most critical factors.
While a larger batch size can improve throughput by increasing compute utilization, it also incurs significant communication overheads.
Thus, a high-bandwidth interconnect is required to fully exploit the benefits of batching in LLM inference~\cite{isca-2025-deepseek}.
Moreover, as the expert distribution becomes more skewed (larger $\Gamma_{imb}^{acc}$), tokens concentrate on a small subset of accelerators, leading to longer communication times.

In summary, reducing the communication cost in MoE requires both high-bandwidth interconnects and an effective mitigation of load imbalance.
The gating operation computes expert scores through a lightweight FC layer, and its computational cost is negligible compared to the expert computation.

\vspace{-0.00in}
\begin{tcolorbox}[boxsep=0pt,left=3pt,right=3pt,arc=0pt]
\emph{\textbf{Observation-8:}}
The interconnect bandwidth and expert load imbalance are the dominant factors that determine the communication time of MoE blocks at large batch sizes.
\end{tcolorbox}
\vspace{-0.00in}

\section{Experimental Setup}
\label{sec:experimental_setup}

To evaluate LLM serving performance in various configurations, we conducted real-system experiments on DGX H100~\cite{nvidia-h100-dgx} and developed an in-house simulator based on LLMSimulator~\cite{github-2025-LLMSimulator, micro-2024-duplex}.
In our simulator, we modeled modern kernel- and system-level optimizations (e.g., FlashAttention~\cite{neurips-2022-flashattention}, FlashMLA~\cite{github-2025-flashmla}, fused kernels, and optimized communication) to ensure fair and realistic execution-time estimation.
We verified the computational characteristics at the node level using a real system.
For inter-node communication time (e.g., dispatch and combine communication in the MoE block), we validated our simulation results against the timing data reported in DeepEP~\cite{github-2025-deepep}.

We configured the accelerator as a modern NVIDIA B200 GPU, whose key parameters are listed in Table~\ref{tab:ai_accelerator}.
By default, we assumed all GPUs in a group are fully connected via NVLink fifth generation, providing 1.8TB/s of bidirectional bandwidth following the NVL72 system topology~\cite{nvidia-gb200-nvl72}.
For each experiment, we specify the number of GPUs per group and note when InfiniBand XDR (100 GB/s) is used for inter-group communication.
We used DeepSeek-R1, Llama4-Maverick, and GPT-3 (key parameters specified in Table~\ref{tab:deepseek_r1_symbol} and Table~\ref{tab:model_configuration_deployment}); all experiments were performed with BF16 precision for all parameters, KV\$, and activations.
We used BF16 as the baseline, but our observations also hold for lower precisions (\eg, FP8), as further discussed in \S\ref{sec:discussion_low_precision}.

To accurately model real-world serving scenarios, we assumed a Zipfian distribution for token routing~\cite{arxiv-2025-fast-alltoall}.
We varied the degree of skewness ($s$) to thoroughly study its impact on system performance.
For better interpretability, we annotated each distribution with the corresponding load imbalance metrics (e.g, $\Gamma_{imb}$) defined in \S\ref{sec:contribution-3}.

Following common practices~\cite{isca-2024-splitwise, osdi-2024-distserve, github-2025-dynamo, isca-2025-windserve}, we assume a \emph{disaggregated} system where the prefill and decode stages are executed on separate machines.
We focus on the decode phase as the prefill phase is generally compute-bound and already achieves high utilization without batching~\cite{osdi-2024-distserve, osdi-2024-sarathi, arxiv-2025-prefillonly}.
Moreover, the insights gained from analyzing the communication time of MoE blocks also apply to the prefill phase as large-batch decode scenarios exhibit similar interconnect traffic patterns to prefill.
For model deployment for the decode system, we set $deg_{\mathrm{TP}}$ to 8 for both GPT-3 and Llama 4-Maverick, chosen to maximize performance of the model, while aligning with the typical 8-GPU topology of NVIDIA DGX systems~\cite{nvidia-dgx-b200}.
For DeepSeek-R1, we set the $deg_{\mathrm{TP}}$ to 1, in accordance with our observation (Obs. 5).

\section{End-to-End Model Execution Analysis}
\label{sec:contribution-4}

\subsection{The Synergistic impact of MLA and MoE}
\label{sec:eval:synergistic}
LLMs adopting MLA and MoE achieve significantly higher throughput than conventional models.
This is because MLA and MoE have a powerful synergistic relationship.
MLA's highly-compressed KV\$ dramatically increases the memory capacity available for batching ($\mathrm{B_{cap}}$).
This, in turn, allows the system to form the large batches required to fully utilize the compute resources of the sparsely activated experts in MoE blocks, which would otherwise be constrained by memory.

Figure~\ref{fig:throughput_latency_32dev} illustrates this by comparing DeepSeek-R1 with Llama4-Maverick and GPT-3.
For a sequence length of 8192, DeepSeek-R1's $\mathrm{B_{cap}}$ (7360) is nearly 60$\times$ larger than GPT-3's (124) and 2.21$\times$ larger than Llama4-Maverick's (3328), which has  an even larger model size.
Consequently, DeepSeek-R1 can be configured with a batch size large enough to approach its ridge point $\mathrm{B_{RP}} (= \! \mathrm{B_{attn}})$, whereas Llama4-Maverick and GPT-3 become memory-capacity-limited long before their compute resources can be saturated.

\begin{figure}[!tb]
\centering
\includegraphics[width=0.99\columnwidth]{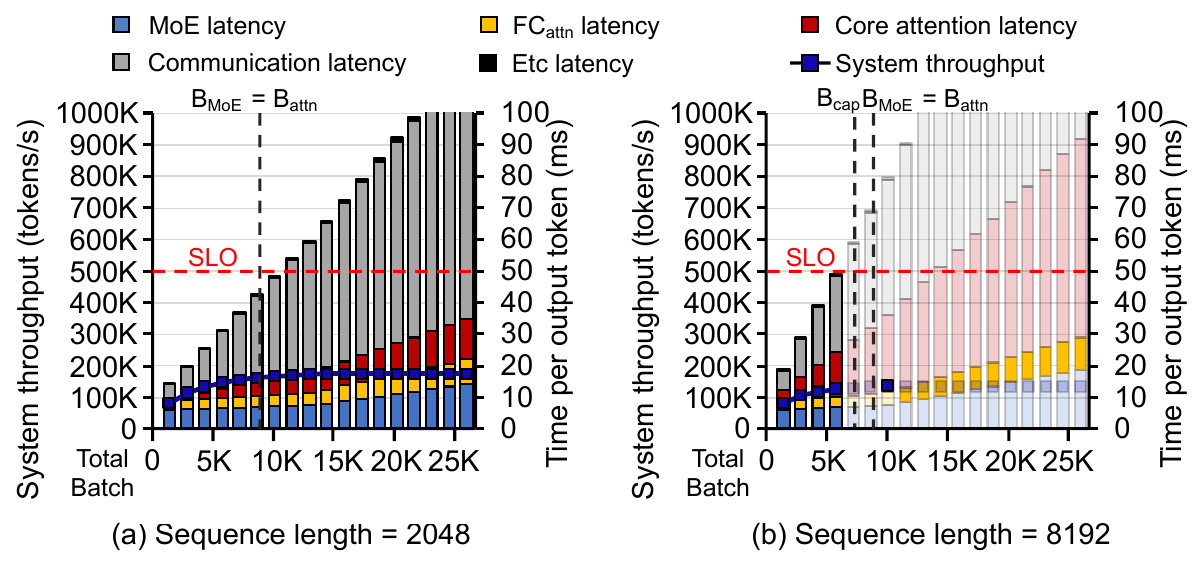}
  \vspace{-0.1in}
  \caption{System throughput and execution time ratio of the decode stage of DeepSeek-R1 when using InfiniBand XDR (100GB/s) among a group of GPUs (DGX), varying sequence lengths and batch sizes. We assume 32 B200 GPU system.}
  \label{fig:low_bw_interconnect}
  \vspace{-0.1in}
\end{figure}

\subsection{The critical role of interconnect}
\label{sec:eval:interconnect}

The performance of a scaled-out MoE-based system is highly sensitive to interconnect bandwidth~\cite{dally-2021-ppin}.
The all-to-all communication pattern, required to dispatch every token to its designated experts and then combine the results, creates dense network traffic that can easily become a bottleneck.
As shown in Figure~\ref{fig:low_bw_interconnect}, moving from a high-bandwidth fabric, such as NVLink, to a lower-bandwidth one, such as InfiniBand, dramatically increases this communication overhead.
For a per-accelerator batch size of 128, our measurements show that a single all-to-all communication task (e.g., dispatch/combine) takes 151.8 µs on a lower-bandwidth fabric, compared with 17.65 µs on the high-bandwidth fabric.
Higher communication latency consumes a larger portion of the per-token time budget, which directly reduces the achievable batch size under a given SLO ($\mathrm{B_{SLO}}$) and leads to underutilization.
Thus, for efficient system deployment, it is critical to have interconnects with high bisection bandwidth.

This sensitivity forces a critical deployment decision: using multiple small, tightly-coupled instances (\eg, \gpus) versus one large, monolithic instance (\eg, \gpub).
Since it is difficult to scale the number of accelerators while maintaining high bisection bandwidth, we vary the interconnect bandwidth of the \gpub configuration to 900 GB/s, \textbf{300 GB/s}, and \textbf{100 GB/s}, which are equal to or lower than that of each \textbf{32 GPU} instance.

\begin{figure}[!tb]
  \center
\includegraphics[width=0.99\columnwidth]{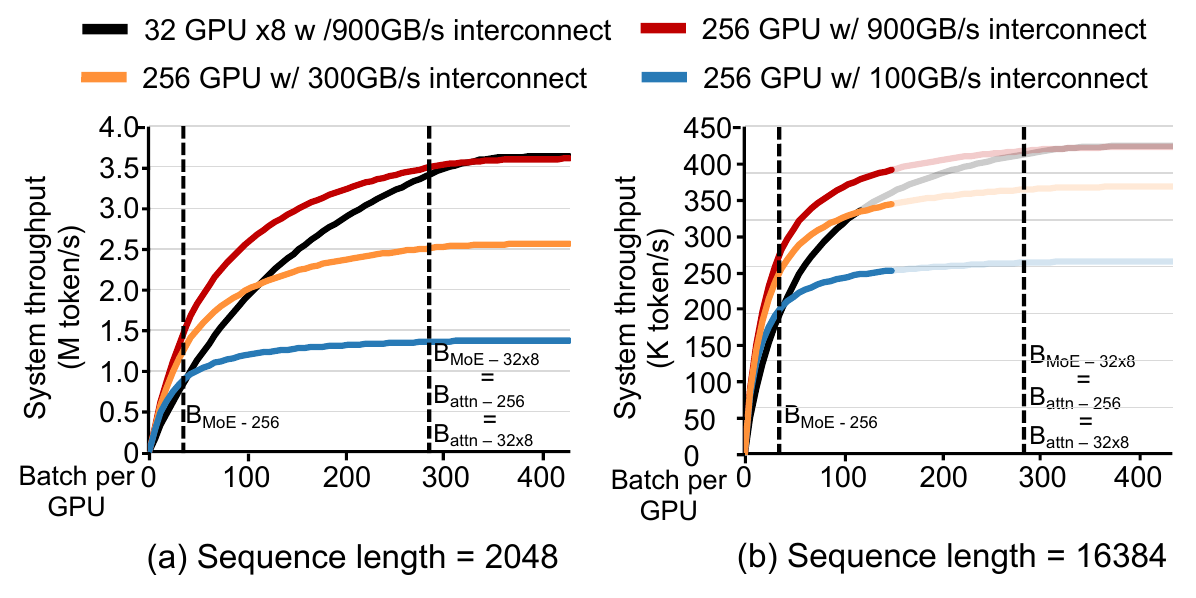}
  \vspace{-0.1in}
  \caption{
  Throughput comparison of \gpub and \gpus systems of the decode stage of DeepSeek-R1 when $L\! =\! 2048$ and $L\! = \! 16384$. 900 GB/s denotes NVLink, while 100 GB/s corresponds to InfiniBand.
  }
  \label{fig:interconnect_bandwidth_32_256}
  \vspace{-0.15in}
\end{figure}

As Figure~\ref{fig:interconnect_bandwidth_32_256} shows, the optimal choice depends on the workload.
For shorter sequences, multiple small instances are more cost-effective because communication is contained within high-bandwidth domains, and the memory overhead of replicating MoE weights is manageable.
When $L\! =\! 2048$ (Figure~\ref{fig:interconnect_bandwidth_32_256}(a)), at a batch size of $\mathrm{B_{RP}}$ to maximize throughput, \gpus achieves equivalent throughput as \gpub with 900 GB/s interconnect bandwidth.
At this point, in \gpub, each GPU is responsible for executing only one expert, but each expert processes 8 times more tokens than in \gpus. 
As the Op/B of experts in \gpub belongs to the compute-bound region, it results in higher latency.
Thus, the latency of MoE blocks becomes similar across the systems. 

For very long sequences (\eg, $L\! =\! 16384$ in Figure~\ref{fig:interconnect_bandwidth_32_256}(b)), however, a single large instance is superior.
The memory savings from storing the massive MoE weights only once by \gpub frees up system-wide capacity for a larger $\mathrm{B_{cap}}$, which is essential for handling the large KV\$, over \gpus.
This leads to higher overall throughput, even if the large-scale interconnect has higher latency.
For example, even with a reduced interconnect bandwidth of 300 GB/s, \gpub delivers better throughput by reducing MoE execution latency.


\subsection{Skewed expert distribution}
\label{sec:eval:expert_skew}

\begin{figure*}[!tb]
  \center
  \includegraphics[width=0.99\textwidth]{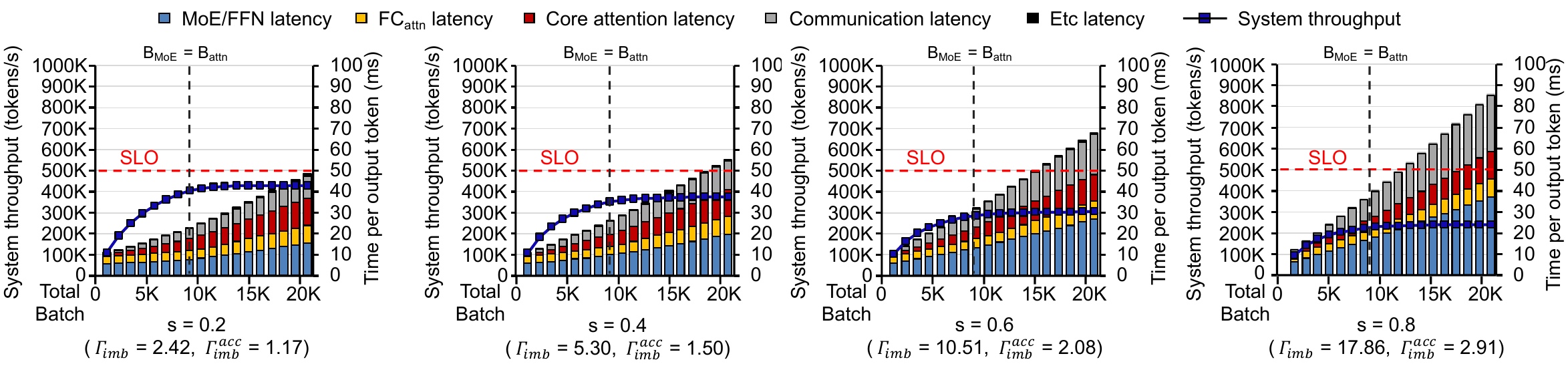}
  \vspace{-0.1in}
  \caption{
  Throughput-latency graph for the decode stages of DeepSeek-R1 with skewed expert routing and 2048 sequence length in 32 GPU system.}
  \label{fig:skewed_throughput_latency_32dev}
  \vspace{-0.15in}
\end{figure*}

Mitigating skewness in expert distribution is essential for achieving high-throughput and low-latency MoE execution.
Figure~\ref{fig:skewed_throughput_latency_32dev} presents the throughput–latency trade-off for the decode stages of DeepSeek-R1 under varying degrees of expert routing skewness ($s$). 
As $s$ increases (from 0.2 to 0.8), the overall system throughput gradually decreases due to the load imbalance among the accelerators.
Also, the latency increases as more tokens are concentrated on a smaller subset of experts. 
As the distribution get more skewed, the rate of increase in both communication latency and MoE latency with respect to the batch size also grows, indicating more severe performance degradation under skewed conditions. 
These results indicate that skewed expert routing reduces the effectiveness of batching, as increasing skewness leads to higher latency and diminishing throughput gains.

\begin{figure}[!tb]
  \center
\includegraphics[width=0.99\columnwidth]{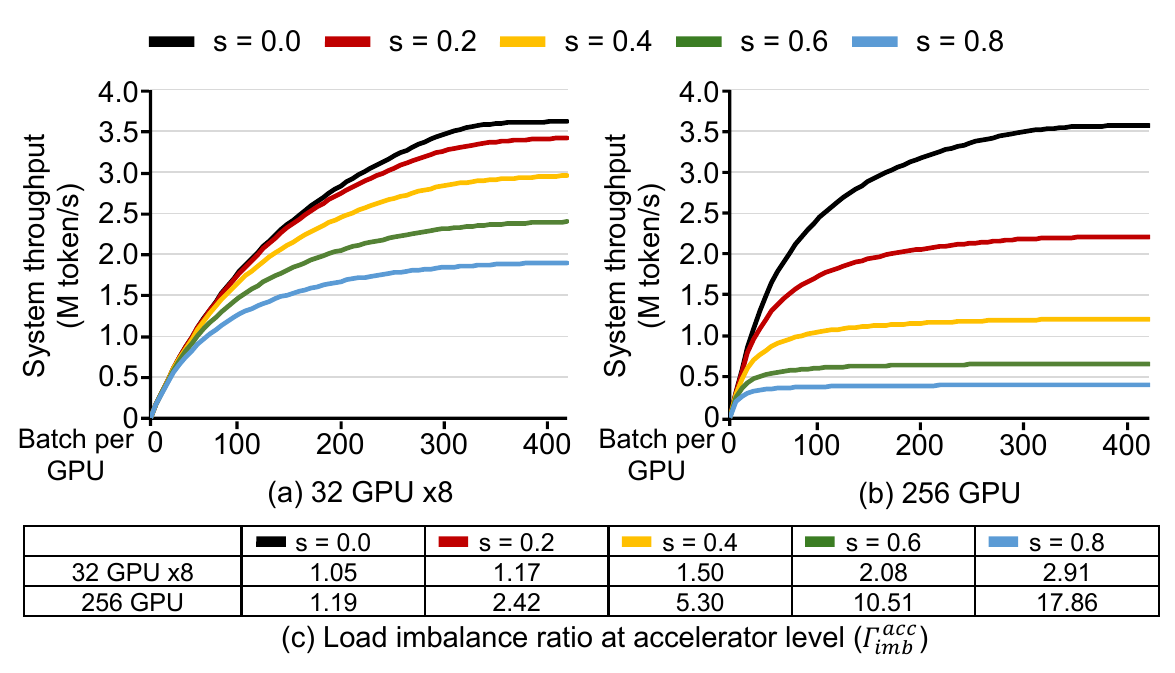}
  \vspace{-0.1in}
  \caption{
  Throughput and load imbalance ratio comparison between \gpub\ and \gpus\ systems for the decode stage of DeepSeek-R1 under varying skewness of expert distribution when $L=2048$. Both systems use a 900~GB/s interconnect among GPUs.
  }
  \label{fig:32x8_256_2048_throughput_skewed}
  \vspace{-0.1in}
\end{figure}

\emph{Our observations remain valid even with a skewed distribution of experts; however, the preferred deployment configurations will be affected by this skewness.} 
Under a uniform random distribution, the batch size that saturates the throughput is close to $\mathrm{B_{MoE}}$. 
However, with skewness, hot experts become saturated before the total batch size reaches $\mathrm{B_{MoE}}$, while cold experts process fewer tokens, resulting in lower \ai\ and reduced throughput. 
When the total batch size increases beyond $\mathrm{B_{MoE}}$, the \ai\ of cold experts can eventually reach $\mathrm{RP_{acc}}$, leading to a larger batch size required for throughput saturation. 
Nevertheless, saturating all experts increases latency, as the hot experts have already reached their maximum throughput and only contribute additional latency without improving overall throughput.
Therefore, service providers must select an appropriate batch size that balances the trade-off between throughput and latency, considering the skewness.

Smaller deployment units such as \gpus can more effectively mitigate the load imbalance compared to a monolithic \gpub.
Figure~\ref{fig:32x8_256_2048_throughput_skewed} compares the system throughput when serving DeepSeek-R1 with different deployment granularities, either using a single deployment of \gpub or eight deployments of \gpus each, under varying levels of expert routing skewness. 
When there is no skew ($s=0.0$), the \gpub configuration achieves higher throughput due to larger aggregate compute and communication bandwidth. 
However, as skewness increases, throughput degrades more severely in \gpub, while \gpus maintains higher throughput.
When the skewness is 0.8, $\Gamma_{imb}^{acc}$ of \gpub is 6.13$\times$ higher than that of \gpus.
In \gpub, each GPU handles only one expert; thus, the token imbalance among the experts directly translates to a load imbalance across GPUs. 
In contrast, in \gpus, each GPU handles 8 experts, which naturally balances the token distribution and mitigates the load imbalance. 
Both systems assume a 900 GB/s interconnect; considering the significantly higher networking cost required to fully connect \gpub, \gpus offers a more balanced and cost-efficient deployment unit for large-scale MoE serving.

\begin{figure}[!tb]
  \center
\includegraphics[width=0.99\columnwidth]{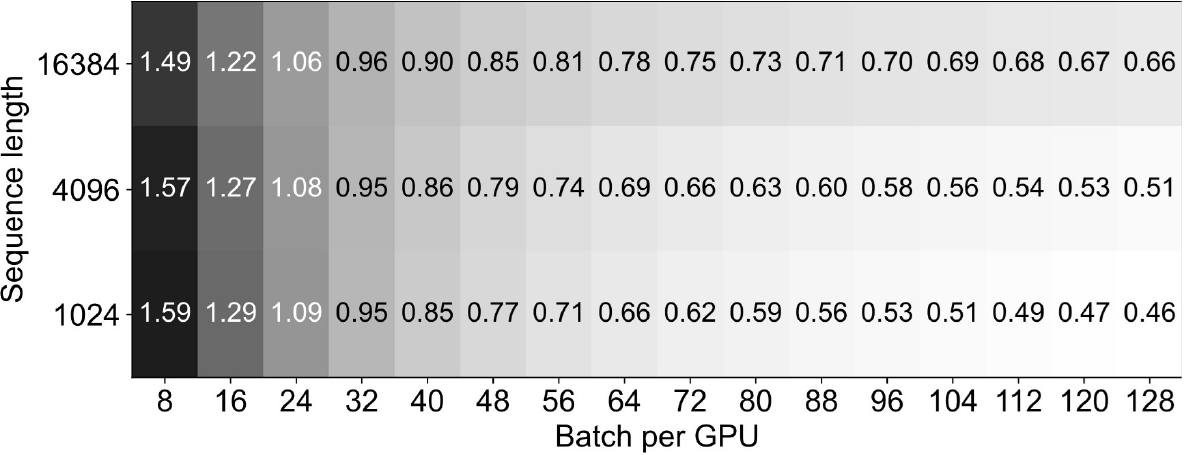}
  \vspace{-0.1in}
  \caption{
  Normalized throughput of Duplex~\cite{micro-2024-duplex}, which PIM devices process only MoE execution, compared to the baseline GPU system. We used PIM devices with $RP_{\mathrm{acc}}$=8, utilizing 4 times of HBM memory bandwidth of GPU.}
  \label{fig:GPUvsPIM}
  \vspace{-0.1in}
\end{figure}

\subsection{Effectiveness of Processing-In-Memory architectures}
At low-batch inference scenarios, where latency-sensitive workloads or on-device inference are required, executing MoE layers on Processing-in-Memory (PIM) architectures~\cite{micro-2024-duplex,asplos-2024-attacc,asplos-2024-neupim,asplos-2025-pim_is_all_you_need, micro-2025-stratum} provides better efficiency by exploiting higher memory bandwidth compared to GPUs.
We modeled Duplex~\cite{micro-2024-duplex}, a state-of-the-art HBM-based PIM architecture designed to accelerate MoE layers, and compared its throughput with that of GPUs.
Figure~\ref{fig:GPUvsPIM} shows normalized throughput improvements when using PIM for MoE execution.

When the batch size per GPU is smaller than 32, PIM devices can effectively reduce latency and increase throughput by processing expert computations faster through their high memory bandwidth.
However, as the batch size increases, PIM devices struggle to sustain performance because the \ai of the experts increases, making computation rather than memory bandwidth the dominant bottleneck.
We conclude that, when MLA and MoE are employed, PIM devices are more suitable for low-batch, low-sequence-length inference scenarios and, in particular, decode stages in such scenarios.

\section{Discussion}
\label{sec:discussion}

\niparagraph{Low weight precision:}
\label{sec:discussion_low_precision}
Recent LLMs support low-precision weights such as FP8 to alleviate memory capacity constraints, while accepting a modest loss in accuracy\cite{arxiv-2024-deepseek-v3,arxiv-2025-deepseek-r1,arxiv-2023-fp8lm,neurips-2022-llm_int_8,icml-2023-smoothquant,iclr-2023-optq, mlsys-2024-awq, arxiv-2024-qqq, mlsys-2025-qserve, mlsys-2024-atom}.
The peak FLOPS of accelerators increase when low-precision weights are used.
For example, latest GPUs 
can achieve up to two times higher peak FLOPS when executing FP8 operations compared to FP16 or BF16.
Thus, $\mathrm{RP_{acc}}$ doubles; however, $\mathrm{B_{RP}}$ remains unchanged because memory access also decreases by half, due to the reduced data size. 
In contrast, $\mathrm{B_{cap}}$ increases because low-precision weights reduce the memory footprint of the model, thereby expanding the available memory capacity for KV\$.
In Figure~\ref{fig:throughput_latency_32dev}(b), when $L\! =\! 8192$, the system is unable to reach $\mathrm{B_{RP}}$ due to the $\mathrm{B_{cap}}$ constraint.
By adopting FP8 for model weights, $\mathrm{B_{cap}}$ increases sufficiently to match $\mathrm{B_{RP}}$, enabling the system to maximize throughput.

\section{Conclusion}
\label{sec:conclusion}

Advances in large language models (LLMs) have reshaped the computational landscape of inference. 
Multi-head Latent Attention (MLA) and Mixture-of-Experts (MoE) move the performance bottleneck away from memory bandwidth. 
With layer reordering, MLA becomes mostly compute-bound, which is well-suited for contemporary accelerators, diminishing the need for dedicated hardware.
MoE achieves scalability through sparse expert activation but demands large batches to sustain utilization; MLA complements this by reducing the KV\$, enabling large-batch inference efficiently even for long sequences.
Finally, we highlight that interconnect bandwidth and expert skewness become the primary factors determining end-to-end performance.
Future serving systems must emphasize high-bandwidth interconnects and balanced workloads to achieve scalable, low-latency LLM serving.


\balance
\bibliographystyle{IEEEtranS}
\bibliography{refs}
\clearpage
\onecolumn
\appendix
\section{Symbol table}
\label{sec:appendix_symbol_table}

\begin{table*}[htbp]
\caption{Symbols used throughout this paper, their descriptions, and the exemplar parameters used in DeepSeek-R1~\cite{arxiv-2025-deepseek-r1}}
\centering
\large
\renewcommand{\arraystretch}{1.3}
\resizebox{\textwidth}{!}{
\begin{tabular}{
l|p{5.7cm}|p{2.6cm}!
{\vrule width 0.6pt\hspace{2.2pt}\vrule width 0.4pt}
l|p{5.7cm}|p{2.6cm}}
\toprule
\textbf{Term} & \textbf{Description} & \textbf{DeepSeek-R1} &
\textbf{Term} & \textbf{Description} & \textbf{DeepSeek-R1} \\
\midrule
TP/DP/EP & Tensor / Data / Expert Parallelism & - & 
\text{$\mathbf{O}_t$} & Context output & - \\
\hline
\text{{$\mathrm{deg_{TP}/_{DP}/_{EP}}$}} & TP / DP / EP degree & - & 
\text{$\mathbf{U}_t$} & Final attention output / FFN input & - \\
\hline
B & Batch size & - &
\text{$\mathbf{H}_t$} & FFN output / Decoder block input & - \\
\hline
\text{$L$} & Sequence length & - &
$\mathrm{RP_{device}}$ & Ridge point of device & - \\
\hline
\text{$n_{\mathrm{dec}}$} & Decoder blocks & 61 &
Q, K, V & Query, Key, Value & - \\
\hline
\text{$d_{\mathrm{emb}}$} & Embedding dimension & \text{7168} &
\text{$\textbf{W}_Q$, $\mathbf{W}_K$, $\mathbf{W}_V$} & Weight for Q, K, V generation & - \\
\hline
\text{$n_{\mathrm{head}}$} & Number of heads & 128 &
\text{$\mathbf{W}_{\mathrm{attn\_out}}$} & Out projection weight in attention & \text{(16384, 7168)} \\
\hline
\text{$d_{\mathrm{head}}$} & Head dimension & \text{128} &
\text{$\mathbf{W}_{\mathrm{gate}}$, $\mathbf{W}_{\mathrm{up}}$} & Weight for gate/up in FFN & \text{(7168, 18432)} \\
\hline
\text{$d_{\mathrm{dec}}$} & Decompressed Q, KV dimension & 16384 &
$\mathbf{W}_{\mathrm{down}}$ & Weight for down in FFN & \text{(18432, 7168)} \\
\hline
\text{$d_{\mathrm{Qco}}$, $d_{\mathrm{KVco}}$} & Compressed Q, KV dimension& \text{1536, 512}  &
\text{$\mathbf{W}_{\mathrm{route}}$} & MoE route weight & \text{(7168, 256)} \\
\hline
\text{$d_{\mathrm{RoPE}}$} & Rotary PE dimension & 64 &
\text{$\mathbf{W}_{\mathrm{exp_n,\, gate}}$, $\mathbf{W}_{\mathrm{exp_n,\, up}}$, $\mathbf{W}_{\mathrm{exp_n,\, down}}$} & MoE up/down projection weights & \text{(7168, 2048)}, \text{(7168, 2048)}, \text{(2048, 7168)} \\
\hline
\text{$d_{\mathrm{MoE}}$} & MoE dimension & 2048 &
\text{$\mathbf{W}_{\mathrm{CQ}}$, $\mathbf{W}_{\mathrm{CKV}}$} & Q comp / KV compression & \text{(7168, 1536)}, \text{(7168, 512)} \\
\hline
$n_\mathrm{k}$ & Top-k experts & 8 &
$\mathbf{W}_\mathrm{DQ}$ & Q decompression weight & \text{(1536, 16384)} \\
\hline
\text{$n_\mathrm{e}$} & Number of experts & 256 &
$\mathbf{W}_{\mathrm{DK}}$, $\mathbf{W}_{\mathrm{DV}}$ & K, V decompression weights & \text{(512, 16384)} \\
\hline
\text{$\mathbf{Q}_{\mathrm{NoPE}}$} & Query vector (No RoPE) & \text{(1, 16384)}&
\text{$\mathbf{W}_{\mathrm{RQ}}$} & RoPE Q weight & (1536, 8192) \\
\hline
\text{$\mathbf{Q}_{\mathrm{RoPE}}$} & Query after RoPE & \text{(1, 8192)} &
\text{$\mathbf{W}_{\mathrm{RK}}$} & RoPE K weight & (7168, 64) \\
\hline
\text{$\mathbf{K}_{\mathrm{RoPE}}$} & Key vector for RoPE & \text{(1, 64)} &
\text{$\mathbf{C}_Q$} & Latent Q (compressed) & - \\
\hline
\text{$\mathbf{S}_{\mathrm{RoPE}}$} & Score output with RoPE & - &
\text{$\mathbf{C}_\mathrm{KV}$} & Latent KV (compressed) & - \\
\hline
\text{$\mathbf{S}_{\mathrm{NoPE}}$} & Score output without RoPE& - &
\text{$n_{\mathrm{acc}}$} & Number of accelerators & - \\
\hline
\text{$\mathbf{S}_t$} & Final score output &
- & - & - \\
\bottomrule
\end{tabular}
}
\label{tab:gpt_symbol}
\end{table*}

\end{document}